\documentclass[%
 reprint,
 twocolumn,
 amsmath,amssymb,
 aps,
 prc,
 floatfix,
]{revtex4}

\usepackage{graphicx}
\usepackage{dcolumn}
\usepackage{bm}
\usepackage{hyperref}

\graphicspath{{figures/}}

\newcommand{\be}{\begin{equation}}
\newcommand{\ee}{\end{equation}}
\newcommand{\nn}{\nonumber}

\begin{document}

\title{``Smashing more than two'':\\ 
Deuteron production in relativistic heavy ion collisions via stochastic multi-particle reactions}

\author{J. Staudenmaier$^{1,4}$} 
\email{staudenmaier@fias.uni-frankfurt.de}

\author{D. Oliinychenko$^{2}$}
\email{oliiny@uw.edu}

\author{J.M. Torres-Rincon$^{1}$}
\email{torres-rincon@itp.uni-frankfurt.de}

\author{H. Elfner$^{3,1,4,5}$}
\email{h.elfner@gsi.de}

\address{1 Institute for Theoretical Physics, Goethe University, Max-von-Laue-Strasse 1, 60438 Frankfurt am Main, Germany}
\address{2 Institute for Nuclear Theory, University of Washington, Seattle, WA, 98195, USA}
\address{3 GSI Helmholtzzentrum f\"ur Schwerionenforschung, Planckstr. 1, 64291 Darmstadt, Germany}
\address{4 Frankfurt Institute for Advanced Studies, Ruth-Moufang-Strasse 1, 60438 Frankfurt am Main, Germany}
\address{5 Helmholtz Research Academy Hesse for FAIR (HFHF), GSI Helmholtz Center, Campus Frankfurt, Max-von-Laue-Straße 12, 60438 Frankfurt am Main, Germany}

\date{\today}

\begin{abstract}
We study the deuteron production via the deuteron pion and nucleon catalysis reactions, $\pi p n \leftrightarrow \pi d$ and $N p n \leftrightarrow N d$, by employing stochastic multi-particle reactions in the hadronic transport approach \texttt{SMASH} for the first time. This is an improvement compared to previous studies, which introduced an artificial fake resonance $d'$ to simulate these $3 \leftrightarrow 2$ reactions as a chain of $2\leftrightarrow 2$ reactions. The derivation of the stochastic criterion for multi-particle reactions is presented in a comprehensive fashion and its implementation is tested against an analytic expression for the scattering rate and the equilibrating particle yields in box calculations. We then study Au + Au collisions at $\sqrt{s_{\mathrm{NN}}} = 7.7$ GeV, where we find that multi-particle collisions substantially reduce the time required for deuterons to reach partial chemical equilibrium with nucleons. Subsequently, the final yield of $d$ is practically independent from the number of $d$ at particlization, confirming the results of previous studies. The mean transverse momentum and the integrated elliptic flow as a function of centrality are rather insensitive to the exact realization of the $2\leftrightarrow 3$ reactions. 
\end{abstract}

\maketitle


\section{Introduction}
In transport simulations of heavy ion collisions the most commonly implemented reactions are $2\to 2$ elastic and inelastic scattering, $1\to 2$ (and sometimes $1\to$ many) resonance decays and $2\to 1$ resonance formations. These reactions are the most likely ones to occur in the dilute limit, where transport approaches are applicable and are easier to implement. However, in certain cases multi-particle reactions, where more than two particles are incoming or outgoing become important. Such cases can be identified by employing the detailed balance principle.

The detailed balance principle states that in an equilibrated system the rate of each elementary process is equal to the rate of the reverse reaction. A sufficient condition for this is the equality of the squared matrix elements for any forward and reverse reaction. This, in turn, follows from time reversal invariance, which is strictly fulfilled for the strong interaction. One consequence of the detailed balance principle is that if a $1\to n$ decay or $2\to n$ reaction is frequent, then in a system close to equilibrium the reverse reaction is frequent too. Moreover, the equality of matrix elements allows to compute the rate of the reverse reaction. Practical cases, where multi-particle reactions have been demonstrated to play a role are the following:

\begin{itemize}
    \item Deuteron production by $N p n \leftrightarrow N d$ reactions in heavy ion collisions at low energies (Nb+Nb collisions at  projectile energy of 650 MeV/nucleon in case of \cite{Danielewicz:1991dh}).
    \item Anti-baryon production by $B\bar{B} \leftrightarrow 3 \text{ mesons}$ reactions  \cite{Cassing:2001ds,Seifert:2017oyb,Seifert:2018bwl} in heavy ion collisions from $\sqrt{s_{\mathrm{NN}}} = 2$ up to 2760 GeV center of mass frame energy per nucleon pair.
    \item Gluon $gg\leftrightarrow ggg$ reactions in a partonic cascade \cite{Xu:2004mz}.
    \item Reactions reverse to many-body decays, such as $3\pi \to \omega$ need to be present to fulfill the detailed balance principle \cite{Weil:2016fxr}.
\end{itemize}

In this work we focus on the three-body deuteron catalysis reactions involving $\pi$ and $N$: $\pi p n \leftrightarrow \pi d$ and $N p n \leftrightarrow N d$. At lower collision energies baryons dominate at mid-rapidity, therefore nucleons are the most frequent catalysts of deuteron production. At energies above $\sqrt{s_{\mathrm{NN}}} \approx 5$ GeV pion catalysis, i.e. $\pi p n \leftrightarrow \pi d$ reactions, start to dominate.

In contrast to this work, deuteron observables in heavy ion collisions are usually computed either by final-state coalescence from nucleons or by a thermal approach assuming chemical equilibrium of deuterons with hadrons, see \cite{Oliinychenko:2020ply} for an overview of the available models. Microscopic deuteron production by pion catalysis is a recent idea introduced in \cite{Oliinychenko:2018ugs} for central Pb+Pb collisions at LHC energies and further tested for non-central collisions and at lower energies down to $\sqrt{s_{\mathrm{NN}}} = $ 7.7 GeV \cite{Oliinychenko:2018odl, Oliinychenko:2020znl}. In all these cases the catalysis reactions proceed rapidly enough to keep deuterons in relative equilibrium with nucleons. This circumstance explains the ``snowballs in hell'' paradox -- the apparent survival of the light nuclei bound by just few MeV at temperatures of more than hundred MeV. The light nuclei do not survive, instead they are destroyed and created at similar rates. While the pion catalysis approach seems to be successful, it also received criticism \cite{Aichelin:2019tnk}, because the implementation of the $\pi pn \leftrightarrow \pi d$ reaction involved a non-existent intermediate resonance $d'$: the reaction was split into two stages $pn \leftrightarrow d'$, $\pi d' \leftrightarrow \pi d$. This purely technical simplification is alleviated in the present work -- the catalysis reactions are implemented directly as a $3\leftrightarrow 2$ process, without involving an artificial intermediate resonance. 

The difficulty of implementing multi-particle reactions arises from the collision criterion. In transport approaches, the most common criterion whether particles collide is based on the comparison of the distance between the particle and a geometric interpretation of the cross-section \cite{Cugnon:1980zz, Bass:1998ca}. So far no generalization of this criterion from two particle to multi-particle reactions is available. Another possible criterion that is easily generalized to multi-particle reactions is directly derived from the collision integral of the Boltzmann equation and formulates a probability (\emph{stochastic rates}) for a reaction \cite{Cassing:2001ds}. Apart from the advantage of the straight-forward generalization to multi-particle reactions, it is inherently boost-invariant.

This work introduces the stochastic criterion in order to treat multi-particle reactions and its application to the deuteron catalysis reactions for the recently established transport approach \texttt{SMASH}~\cite{Weil:2016zrk}. The new treatment of multi-particle collisions allows to assess the previously found conclusions for the deuteron production on a more solid basis. It also allows to investigate the difference arising from modelling multi-particle reactions as a chain of two-body reactions (e.g. $\pi\pi\pi\rightarrow\rho\pi\rightarrow\omega$). This helper construct with intermediate resonances is employed in \texttt{SMASH} in several places to adhere to geometric criterion and maintain detailed balance.

Employing a probabilistic collision criterion to enable direct multi-particle reactions is already explored in the literature, starting with \cite{LANG1993391, Danielewicz:1991dh}. While \cite{LANG1993391} focused on the optimisation of computing time and only applied the method for two-body reactions, the authors in \cite{Danielewicz:1991dh} also discussed the production of deuterons. The approach presented here and the one from \cite{Danielewicz:1991dh} both include the deuteron catalysis reaction involving nucleons, $Npn\rightarrow Nd$. For the low energies discussed in \cite{Danielewicz:1991dh} this reaction is likely sufficient, since the system is dominated by $N$, but for the higher (intermediate) beam energy discussed here the pion catalysis reaction, $\pi pn\rightarrow \pi d$, has to be included as well~\cite{Oliinychenko:2020znl}. This is similar to a recent study for high beam energies \cite{Sun:2021dlz}.
Other authors focus on the $B\bar{B}$ annihilation reactions that produce multiple mesons \cite{Cassing:2001ds, Seifert:2017oyb, Seifert:2018bwl} with the PHSD (Parton-Hadron String Dynamics) approach \cite{Bratkovskaya:2011wp}. On the parton level, \emph{stochastic rates} are studied with the parton transport approach BAMPS (Boltzmann Approach of Multi-Parton Scatterings) including the $ggg\leftrightarrow gg$ gluon bremsstrahlung reactions \cite{Xu:2004mz}. The hadronic transport approach GiBUU also allows to employ a probabilistic collision criterion \cite{Buss:2011mx}.

The structure of this article is as follows: First the employed transport approach and the treatment of deuterons is described in Sec.~\ref{sec:model}. After the description of the geometric collision criterion so far employed in \texttt{SMASH}, the stochastic collision criterion is introduced  by providing a sketch of its derivation. The stochastic criterion is validated by applying it to two-body reactions. Then, the treatment of multi-particle deuteron reactions is discussed. The stochastic multi-particle reactions are again validated by comparing the scattering rates to analytic results obtained from rate equations in Sec.~\ref{sec:rates}. The main results for the production of deuterons in Au+Au collision are presented in Sec.~\ref{sec:results} and Sec.~\ref{sec:conclusions} closes with a summary and conclusions. 

\section{Model Description \label{sec:model}}

\subsection{Transport Approach: \texttt{SMASH}}
The hadronic transport approach employed for the present study is \texttt{SMASH}~\cite{Weil:2016zrk}, which is extended to include a stochastic collision criterion (see Sec.~\ref{sec:coll-crit}) and multi-particle reactions (see Sec.~\ref{sec:multi-part}) that obey the detailed balance principle.

\texttt{SMASH} includes most hadronic degrees of freedom listed by the Particle Data Group~\cite{Zyla:2020zbs} up to a mass of $2.3$ GeV. See~\cite{Weil:2016zrk} for the list of included degrees of freedom and \cite{Steinberg:2018jvv} for an update focused on the strange sector. The transport approach constitutes an effective solution of the Boltzmann equation by mapping the collision term to binary elastic and inelastic scatterings as well as the formation and decay of excited resonances. Additionally, string fragmentation to describe the high-energy cross sections and baryon-antibaryon annihilation together with the mentioned newly-introduced multi-particle reactions (see Sec.~\ref{sec:multi-part}) are included. The reader is referred to~\cite{Mohs:2019iee} for details on the string fragmentation treatment. Resonances have vacuum properties and are tuned to reproduce the elementary cross sections. The Breit-Wigner spectral functions include a mass-dependent width, which follows the idea from Manley and Saleski in~\cite{Manley:1992yb} (but with updated resonance properties). The approach is able to perform infinite-matter (\emph{box}), expanding sphere, collider and afterburner (\emph{list}) calculations.

The version used throughout this work is \texttt{SMASH-2.0.1}~\cite{dmytro_oliinychenko_2020_4336358}. This new major version introduces the here discussed multi-particle reactions together with an integration of a hydrodynamics phase and improvements for high-energy AA collisions, when Pythia is required for hard scatterings. Also, the distance definition for the geometric criterion is updated to a fully covariant formulation~\cite{Hirano:2012yy}.

\subsection{Deuterons in \texttt{SMASH}}

Deuterons in \texttt{SMASH}~\cite{Oliinychenko:2018ugs, Oliinychenko:2020znl} are treated like on-shell point-particles as in~\cite{Danielewicz:1991dh, Oh:2009gx,Longacre:2013apa}. The applicability of this approach is not strictly justified for the whole time of the evolution: the deuteron mean free path at the start of the afterburner evolution is comparable to twice the geometric size of deuteron wavefunction, although at later stages the mean free path increases making the approach more justifiable. Because of this we should strictly speaking call our ``deuterons'' correlated nucleon pairs. However, it seems that the inclusion of the finite deuteron size is not important in Pb+Pb collisions -- a recent study  \cite{Sun:2021dlz} on the inclusion of the finite deuteron size in a similar approach only found a significant effect for a much smaller fireball created in $pp$ collisions, which are not discussed in this work. Our approach with on-shell point-like $d$ describes the data for heavy-ion collisions reasonably well, and is able to capture the fact that deuterons (or ``correlated nucleon pairs'') are created and destroyed.  In our model the formation and disintegration of deuteron is catalyzed by pions or nucleons in the following reactions and their CPT-conjugates: $\pi d \leftrightarrow \pi n p$, $N d \leftrightarrow N n p$, $\bar{N} d \leftrightarrow \bar{N} n p$ and $\pi d \leftrightarrow N N$. Additionally, elastic reactions for $\pi d$, $N d$ and $\bar{N} d$ are included.

The challenge in a microscopic transport approach is to treat the above mentioned $X d \leftrightarrow X n p$ ($X=\{\pi,N,\bar{N}\}$) catalysis reactions: Usually a geometric collision criterion is employed and it is not clear how the $3 \rightarrow 2$ reaction should be treated, since no generalization of a geometric collision criterion for more than 2 particles is available. Therefore, the $3 \leftrightarrow 2$ is broken down into two steps with a fake decaying resonance (called $d'$) in the intermediate step, so that the reaction (chain) only contains 2-body collisions: $X d \leftrightarrow X d' \leftrightarrow X n p$. The properties of the $d'$ are chosen in order to reproduce the $\pi d$, $N d$ and $\bar{N} d$ cross section and to achieve a lifetime that lasts the time $n$ and $p$ spend flying by each other. This treatment of the deuteron reactions is introduced in detail in~\cite{Oliinychenko:2018ugs} and will be labeled in the following as the \emph{2-to-2 stochastic} or \emph{2-to-2 geometric} treatment depending on the employed collision criterion.

The presented work introduces an alternative to the above described approach in the following, by employing a stochastic collision criterion, which allows for a direct (one-step) treatment of $3 \leftrightarrow 2$ reactions (labeled as \emph{3-to-2 stochastic}, see Sec.~\ref{sec:multi-part}).

\subsection{Collision Criteria \label{sec:coll-crit}}

At the core of microscopic (transport) approaches like \texttt{SMASH}, where point-like particles propagate and interact, the decision if and when they collide has to be made. Two different categories are known for this decision: Geometric and stochastic collision criteria.

\subsubsection{Geometric Criterion \label{sec:geo-crit}}

The most common approach to decide when two particles collide is based on a geometric interpretation of the cross section. The transverse distance of closest approach $d_T$ has to smaller than the interaction distance $d_{int}$ given by the cross section:

\begin{equation}
    d_T < d_{int} = \sqrt{\frac{\sigma}{\pi}} \ .
\end{equation}

There are several definitions available for the transverse distance of closest approach $d_T$ of two particles~\cite{Bass:1998ca, Kodama:1983yk, Hirano:2012yy}. \texttt{SMASH} followed the approach introduced by UrQMD~\cite{Bass:1998ca} up to version \texttt{SMASH-2.0}, where the approach is changed to the fully covariant formulation given in~\cite{Hirano:2012yy}. One main disadvantage of the criterion is the lack of a straight-forward generalization of $d_T$ to more than two particles.

\subsubsection{Stochastic Criterion} \label{sec:st-crit}

In contrast to the geometric criterion, the stochastic criterion defines a probability for a reaction of a given particle set. The probability is defined as the number of reactions $\Delta N_{\textrm{reactions}}$ over the number of all possible particle combinations inside a sub-volume $\Delta^3 x$ and time interval $\Delta t$, 
\be 
P_{n \rightarrow m} = \frac{\Delta N_{\textrm{reactions}}}{\prod_{j=1}^n \Delta N_j} \ . \label{eq:Pnmsimp}
\ee
The final expression of $P_{n\rightarrow m}$ is worked out in App.~\ref{sec:app-pnm}. 

The stochastic criterion is inherently boost-invariant. This idea was already explored in the literature before in~\cite{LANG1993391, Danielewicz:1991dh,Cassing:2001ds,Xu:2004mz,Buss:2011mx,Seifert:2017oyb, Seifert:2018bwl}, also often referred to as the \emph{local-ensemble method}. 

Several special cases (for $n,m=1,2,3$) are presented in App.~\ref{sec:app-pnm}. In particular for a decay process $n=1$,
\begin{equation} \label{eq:p12}
P_{1 \rightarrow m} = \Delta t \frac{M}{E_1} \Gamma_{1 \rightarrow m} (M) \ ,
\end{equation} 
where $m$ is the number of final particles after the decay, and $M$ is the total energy in the center-of-mass of the system (or the mass of the initial resonance in its rest frame). This decay probability is the same already used in \texttt{SMASH} (cf. Eq.~(38) in~\cite{Weil:2016zrk}).

For $n=2$ one can write the probability in terms of the total cross section,
\begin{equation}
P_{2 \rightarrow m} = \frac{\Delta t}{\Delta^3 x} v_{\textrm{rel}} \sigma_{2\rightarrow n} (M) \ ,     
\end{equation}
where the relative velocity is defined in Eq.~(\ref{eq:vrel}). For $m=1$ the cross section is given in Eq.~(\ref{eq:xsec1}), and coincides with $\sigma_{2\rightarrow 1}$ used in~\cite{Weil:2016zrk}. In this paper we use the case $m=3$, and $\sigma_{2\rightarrow 3}$ is fitted to available experimental data in~\cite{Oliinychenko:2018ugs} as described above.

Other cases can be worked out along the lines given in App.~\ref{sec:app-pnm}. Even though, this straight-forward generalization to arbitrary $n \rightarrow m$ reactions is a large advantage compared to geometric criteria, Eq.~(\ref{eq:Pnmbig}) cannot be applied directly, since the matrix element is generally unknown.

When employing the test-particle method the particle number is scaled with the test-particle number $\Delta N_{i} \rightarrow \Delta N_{i} N_{\textrm{test}}$ in the here presented notation. The number of collision scales similar: $\Delta N_{\textrm{reactions}}^{n\rightarrow m} \rightarrow\Delta N_{\textrm{reactions}}^{n\rightarrow m} N_{\textrm{test}}$. Therefore the scaled probability $P'_{n \rightarrow m}$ when employing the test-particle method is

\begin{equation} \label{eq:ptp}
P'_{n \rightarrow m} = \frac{P_{n \rightarrow m}}{N_{\textrm{test}}^{n-1}}.
\end{equation}

For the numerical realization of the stochastic criterion, the space is divided into equally sized cells ($\Delta^3x$) and the probability is calculated for all 2-particle and 3-particle combinations within each time-step, so only particles within a cell are able to interact. The calculated probability is used for a Monte-Carlo decision i.e. if a generated random number between $0$ and $1$ is smaller or equal than the probability the reaction is accepted. The collision time is randomly chosen within the given time-step $\Delta t$. 

In contrast to the geometric criterion, the stochastic criterion is a strictly time-step based method. The time-steps, therefore, have to be chosen small enough that the assumption that each particle only interacts once per time-step is justified and the defined probability is not exceeding $1$ for a given $\Delta^3x$. Also the cells $\Delta^3x$ have to be chosen sufficiently small, since only in the limit of $\Delta^3x\rightarrow 0$ (and $\Delta t\rightarrow 0$) the numerical solution matches the exact solution of the Boltzmann equation. At the same time, the cells still have to be sufficiently filled with particles.

\begin{figure}[htb]
    \centering
    \includegraphics[width=0.45\textwidth]{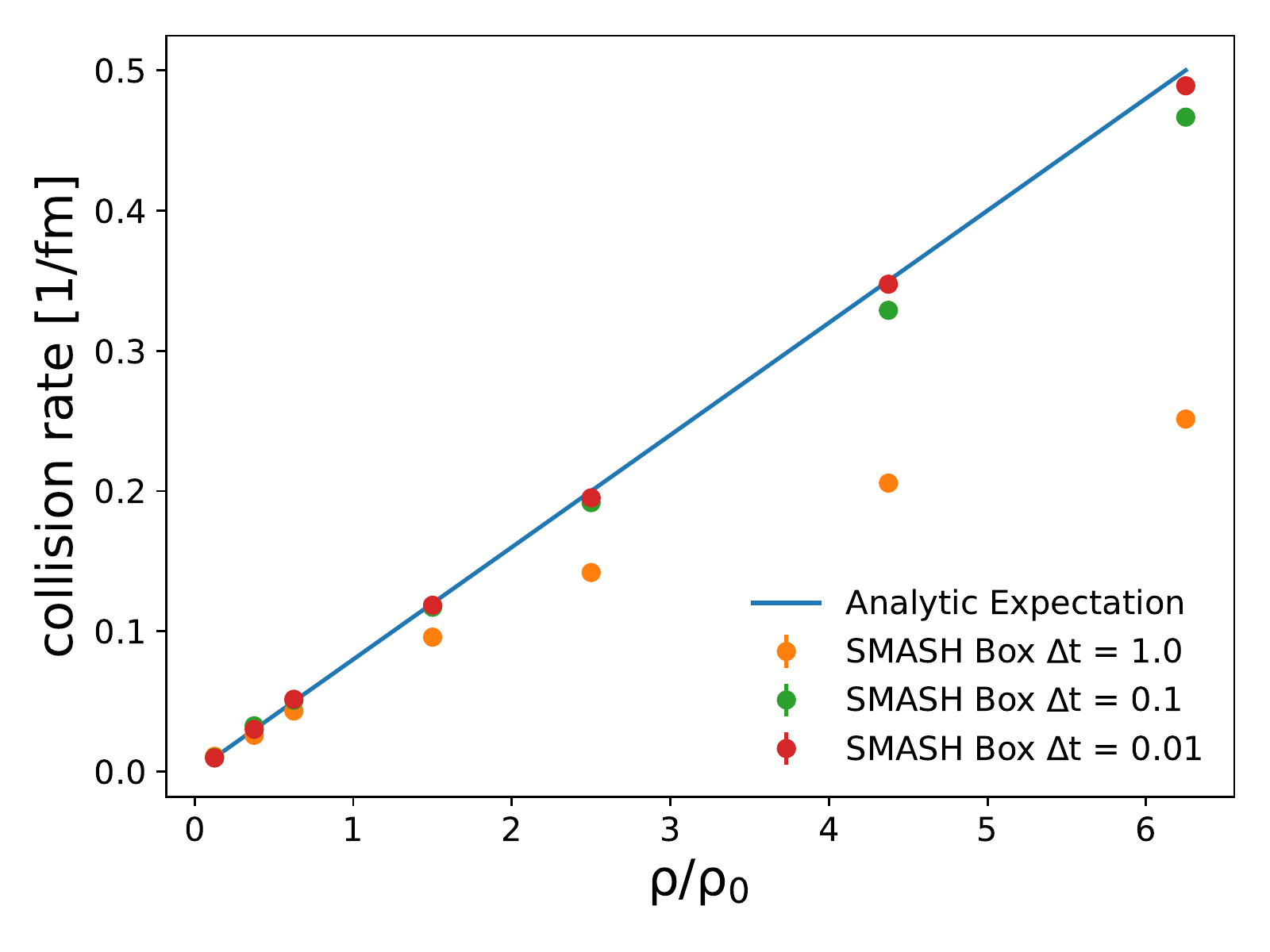}
    \caption{Collision rate in a box with $\pi^0$ mesons interacting with a constant elastic cross section of $\sigma = 10$ mb for different densities and time-step sizes compared with the analytic expectation of $\rho\sigma$. Density given in units of $\rho_0=0.16 \, \rm{fm^{-3}}$.}
    \label{fig:coll-rate}
\end{figure}

To verify the stochastic collision criterion, first the collision rate is studied in Fig.~\ref{fig:coll-rate}. The expected collision rate for the case of a box filled with $\pi^0$ that interact via a constant elastic cross section is given by $\rho \sigma$. Fig.~\ref{fig:coll-rate} shows that the criterion performs well, if the time-step is chosen small enough as mentioned above. Otherwise it exhibits the usual limitation of time-step based approaches: When the density is large and the time-step is also large more than one interaction per time-step is expected and therefore the collision rate is underestimated. It is furthermore checked that the scattering rate is consistently reproducing the analytic expectation  also when varying other parameters like the box size, the test-particle number, the cross section and the temperature of the box. 

\begin{figure}[htb]
    \centering
    \includegraphics[width=0.45\textwidth]{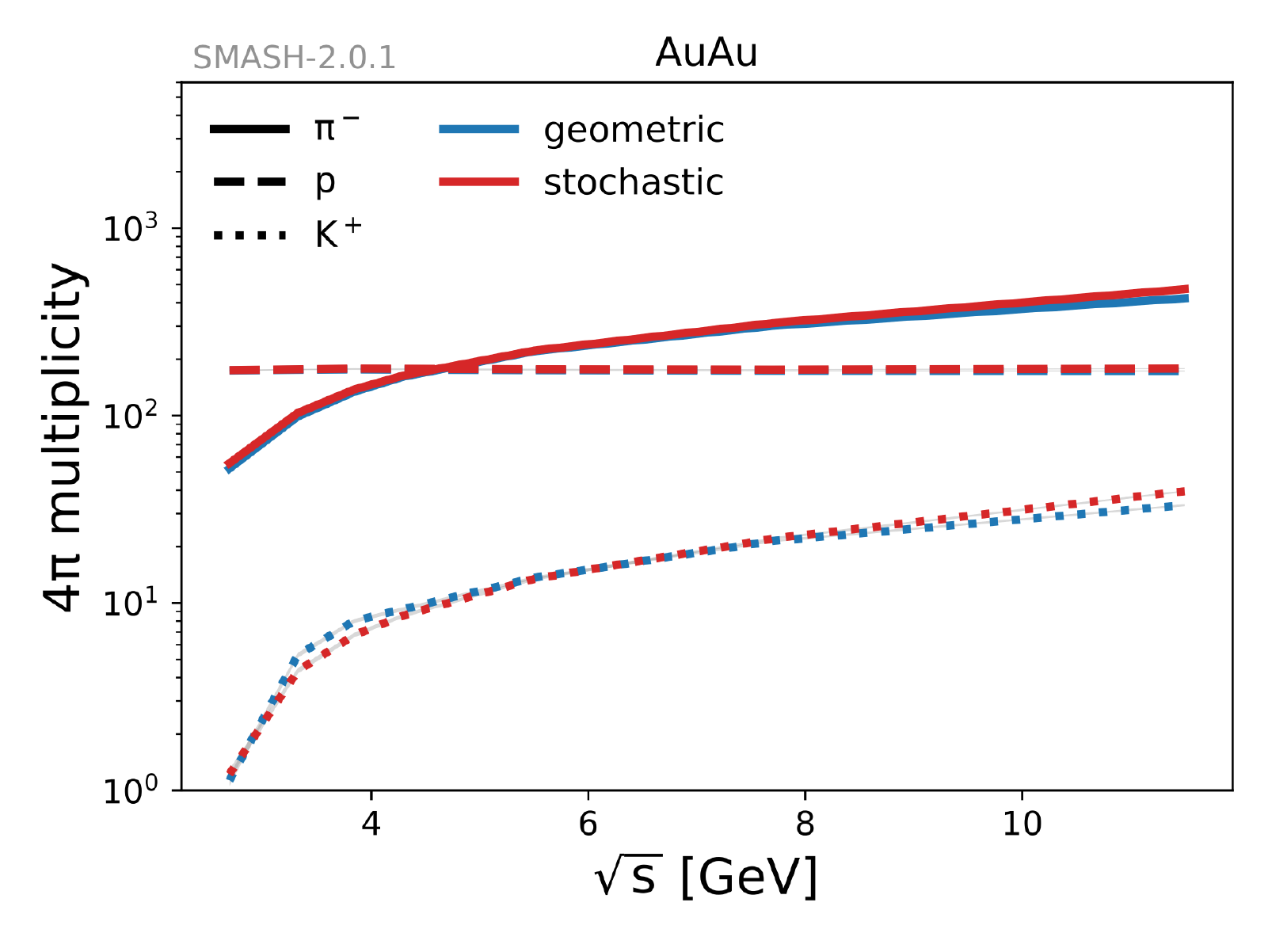}
    \caption{Particle multiplicities ($\pi$, $K$ and $p$) in AuAu collisions with the geometric and stochastic collision criterion for different collision energies.}
    \label{fig:e-scan}
\end{figure}

Complementing the result for the collision rate in the box, produced particle multiplicities from heavy-ion reactions are shown in Fig.~\ref{fig:e-scan} for different beam energies. Since the collision criterion is in the end an implementation detail, it is reassuring that the particle yield does not depend on the chosen collision criterion as Fig.~\ref{fig:e-scan} exemplifies. The stochastic collision criterion matches the previous results with the geometric criterion for a set of abundant hadronic species as a function of beam energy.

\subsection{Multi-particle Reactions \label{sec:multi-part}}

The main class of multi-particle reactions realized with the stochastic collision criterion are $3\leftrightarrow 2$ reactions. These are needed for the description of creation and annihilation of deuterons in the hadron gas. Expressions for $P_{2\rightarrow3}$ and $P_{3\rightarrow 2}$ are derived in Eqs.~(\ref{eq:P23}) and (\ref{eq:P32}), respectively. For completeness, similar calculations for $3\leftrightarrow 1$ reactions are given in Appendix~\ref{sec:app-31}.

Since the scattering matrix element is commonly unknown, Eq.~(\ref{eq:Pnmbig}) cannot be directly used to calculate the required probability. Assuming that the scattering matrix element is independent of the final momenta, the probability for the reverse process of a decay ($n \rightarrow 1$) or a 2-body scatterings ($n \rightarrow 2$) can be expressed in terms of the known decay width or cross section of the $\{1,2\} \rightarrow n$ process. This is the case, because the scattering amplitude is the same for the forward and backward process (\emph{time-reversal invariance}).

Let us focus here on the $P_{3 \rightarrow2}$ probability applied to the deuteron case. The stochastic criterion allows for a direct (one-step) treatment of $3 \rightarrow 2$ reactions, which is applied to the deuteron formation catalysis reactions $\pi p n \leftrightarrow \pi d$, $N p n \leftrightarrow N d$ and $\bar{N} p n \leftrightarrow \bar{N} d$ (and labeled as the \emph{3-to-2 stochastic} treatment in the results below.). 

The probability for a 3-to-2 process is given as

\begin{align} 
P_{3\rightarrow 2} &  =\left( \frac{g_{1'} g_{2'}}{g_1 g_2 g_3} \right) \frac{\cal{S}!}{\cal{S}'!} \frac{\Delta t}{(\Delta^3 x)^2} \frac{E_{1'}E_{2'}}{2E_1E_2E_3} \frac{\Phi_2(s)}{\Phi_3(s)} \nonumber \\ 
&\times v_{\textrm{rel}} \sigma_{2\rightarrow 3}(\sqrt{s}) \ ,
\end{align}
where $\Phi_2,\Phi_3$ are the 2- and 3-body phase space corresponding to final and initial states, respectively (see also (\ref{eq:P32})). 

The symmetry factors for the reactions $\pi pn \rightarrow \pi d$, $\bar{N}pn \rightarrow \bar{N}d$ and their charge conjugated are $\frac{\cal{S}!}{\cal{S}'!} = 1$, 
while for $Npn \rightarrow Nd$ and it charge conjugated, as there are 2 identical particles in the initial state, are $\frac{\cal{S}!}{\cal{S}'!} = 2 $ .
The spin degeneracies read
\be g_{d}/3=g_{N,p,n}/2=g_\pi=1 \ . \ee

Note that the derivation presented in the Appendix~\ref{sec:app-pnm} resembles the one presented in \cite{Cassing:2001ds} and \cite{Xu:2004mz}. The absence of a known scattering matrix element makes additional steps necessary in the probability derivation compared to \cite{Xu:2004mz}. A similar idea as in \cite{Cassing:2001ds} is therefore followed, where the scattering matrix element is assumed to be only dependent on the initial center-of-mass energy. This approach differs from the idea in \cite{Danielewicz:1991dh}, where the matrix element is factorized into two terms as an approximation (also discussed in \cite{Xu:2004mz}). This factorization results in a different formula for the 3-to-2 collision probability containing the two-body $NN$ cross section and a momentum dependent volume \cite{Danielewicz:1991dh}. The approach presented here, while also making an assumption about the matrix element, is more general in the sense that it allows to treat all processes, where a decay width or cross section for the reverse reaction is known.

\subsection{Hybrid Approach: Hydrodynamics and Cooper-Frye}

In this work we focus on the deuteron production in the hadronic afterburner. Therefore,
we describe modelling of the earlier stages of the heavy-ion collisions only briefly.
The expansion and cooling of the fireball during the dense stage is simulated by the
relativistic 3+1-dimensional open-source hydrodynamic code \texttt{MUSIC v3.0} \cite{Schenke:2010nt,Schenke:2011bn,Paquet:2015lta,Denicol:2018wdp}. The initial condition for \texttt{MUSIC} is a smooth parametrized energy density and baryon density as functions of spatial coordinates. The parametrization is described in \cite{Shen:2020jwv}, it is tuned to reproduce charged particle rapidity distribution, pion midrapidity yields, and net-proton midrapidity yields. The initial energy-momentum tensor is assumed to have an ideal fluid form $T^{\mu\nu} = (\epsilon + p)u^{\mu}u^{\nu} - p g^{\mu\nu}$. The flow at initial eigen-time $\tau_0$ ($\tau_0 = 3.6$ fm/c at $\sqrt{s_{\mathrm{NN}}} = $7.7 GeV) is assumed to be only longitudinal and have Bjorken form $u^\mu = (\cosh \eta_s, 0, 0, \sinh \eta_s)$. The equation of state combined with hydrodynamic equations is a lattice QCD based ``NEOS-BSQ'' equation of state $p=p(\epsilon, n_B)$ described in Ref.~\cite{Monnai:2019hkn}. Shear viscous corrections are included with a specific shear viscosity $\eta T/(e + P) = 0.1$, while bulk viscous corrections and baryon number diffusion are neglected. Particlization is performed at a constant energy-density hypersurface, $\epsilon(\tau,x,y,\eta_s) = 0.26$ GeV/fm$^3$. This corresponds to a line in $(T,\mu)$ plane; the distribution of hypersurface cells number at $|\eta_s| < 0.5$ peaks at $(T,\mu) = (141 \pm 4, 340 \pm 50)$ MeV, where the numbers after $\pm$ denote an approximate width of the peak. This is slightly lower energy density and baryon chemical potential, but almost the same temperature compared to the chemical freeze-out in the thermal model fit of the hadron yields in central AuAu collisions at $\sqrt{s_{\mathrm{NN}}} = 7.7$ GeV, where $(T, \mu) = (143.8 \pm 2.7, 399.8 \pm 13.3)$ MeV \cite{Adamczyk:2017iwn}. The particlization procedure is a standard grand-canonical Cooper-Frye particlization with a Grad 14-moments ansatz for shear viscous corrections, see \cite{Schenke:2010nt, Schenke:2011bn} for details. Particles from particlization are inserted into hadronic afterburner and rolled back to equal time, but are forbidden to interact until their actual emergence time. Deuterons may be sampled or not at the particlization, this is mentioned specifically in the text. The $d'$ is never sampled.

Overall, the most important features of the hydrodynamic part of our simulation is that it provides
a reasonable space-time distribution of nucleons and pions at particlization. The nucleon and pion rapidity and transverse momentum spectra reproduce experimental data rather well. Based on our previous work, these are sufficient prerequisites to describe the measured deuteron spectra
by introducing deuteron-creating reactions into the afterburner.

\section{Validation of stochastic multi-particle rates in the box \label{sec:rates}}

Before showing the physics results, let us start by demonstrating that the implementation of the stochastic criterion and the newly introduced 3-to-2 multi-particle reactions work as expected. For this purpose, a particle configuration is initialized in a box with periodic boundary conditions and compared to the analytic expectation. The goal is to ensure that the content of the box equilibrates, that it equilibrates to the correct state, and that particle multiplicities change at the expected rate in the process of equilibration. This method of testing has proven very useful, because it checks detailed balance and reaction rates, and is sensitive even to minor errors in implementation.

\subsection{Rate equations}

The analytic expectation is provided by rate equations of the same form as introduced in \cite{Pan:2014caa} with only one exception -- we take the spectral function into account, when we compute thermally averaged resonance widths $\langle \Gamma \rangle$. The necessary derivation and notation for writing the rate equations for the specific reaction systems is given in Appendix~\ref{sec:app-rate-equations}. Below, only the resulting systems of rate equations are given.

The first case is the $d$, $d'$, $\pi$, $N$ system with $2\leftrightarrow 2$ reactions, where in addition to elastic collisions the following reactions are allowed:

\begin{align}
    \begin{cases}
			p n \leftrightarrow d'\\
            \pi d' \leftrightarrow \pi d \\
            N d' \leftrightarrow N d
    \end{cases}
\end{align}

Denoting time derivative $\frac{d\lambda}{dt}$ as $\dot{\lambda}$ the system of rate equations for this case is given as,

\begin{align} \label{eq:piNddprime_rates}
    \begin{cases}
		n_d^{th} \dot{\lambda_d} &= (R_{\pi d} + R_{N d}) (\lambda_{d'} - \lambda_d) \\
		n_{d'}^{th} \dot{\lambda_{d'}} &= - (R_{\pi d} + R_{N d}) (\lambda_{d'} - \lambda_d) + R_{d'} (\lambda_p^2 - \lambda_{d'}) \\
		n_{p}^{th} \dot{\lambda_{p}} &= - R_{d'} (\lambda_p^2 - \lambda_{d'}) \\
		\dot{\lambda_{\pi}} &= 0 \\
	    R_{\pi d} &= \langle \sigma v_{\textrm{rel}}\rangle_{\pi d} \ n_{\pi}^{th} n_d^{th} \lambda_{\pi}	\\
	    R_{N d} &= 2 \langle \sigma v_{\textrm{rel}}\rangle_{N d} \ n_{p}^{th} n_d^{th} \lambda_p	\\
	    R_{d'} &= \langle \Gamma \rangle_{d'} \ n_{d'}^{th} \ .
    \end{cases}
\end{align}

Here the amount of protons and neutrons is assumed equal. The initial conditions are determined by the content of the box at time $t = 0$. The thermally averaged width of $d'$, $\langle \Gamma_{d'}\rangle$, was artificially divided by factor 2 to agree with the simulation in Fig. \ref{fig:box_rates}. The need in factor 2 might emerge from the fact that the $d'$ spectral function has a long high-mass tail, which takes more time to equilibrate than expected.

The second case is a system of $d$, $\pi$, $N$ with $3\leftrightarrow 2$ reactions, where the following reactions are allowed:

\begin{align}
    \begin{cases}
            \pi p n \leftrightarrow \pi d \\
            N p n \leftrightarrow N d
    \end{cases}
\end{align}

Resulting in the corresponding rate equations as follows,

\begin{align}
    \begin{cases}
		n_d^{th} \dot{\lambda_d} &= (R_{\pi d} + R_{N d}) (\lambda_{p}^2 - \lambda_d) \\
		n_{p}^{th} \dot{\lambda_{N}} &= - (R_{\pi d} + R_{N d}) (\lambda_{p}^2 - \lambda_d) \\
		\dot{\lambda_{\pi}} &= 0 \\
	    R_{\pi d} &= \langle \sigma v_{\textrm{rel}}\rangle_{\pi d} \ n_{\pi}^{th} n_d^{th} \lambda_{\pi}	\\
	    R_{N d} &= 2 \langle \sigma v_{\textrm{rel}}\rangle_{N d} \ n_{p}^{th} n_d^{th} \lambda_N \ .
    \end{cases}
\end{align}

In this second case, no additional correction factors need to be applied, since it avoids the fictitious $d'$ resonance in the first place.

\subsection{Comparing analytic results to simulations}

Let us compare the solutions of the rate equations above to corresponding simulations in a box, which allows to probe the equilibrium properties of the employed reaction treatment and collision criteria. The verification of the approach in a static box scenario is the basis for further exploration of dynamic non-equilibrium systems like nucleus-nucleus collisions in the remainder of this work.
\begin{figure}[htb]
    \centering
    \includegraphics[width=\columnwidth]{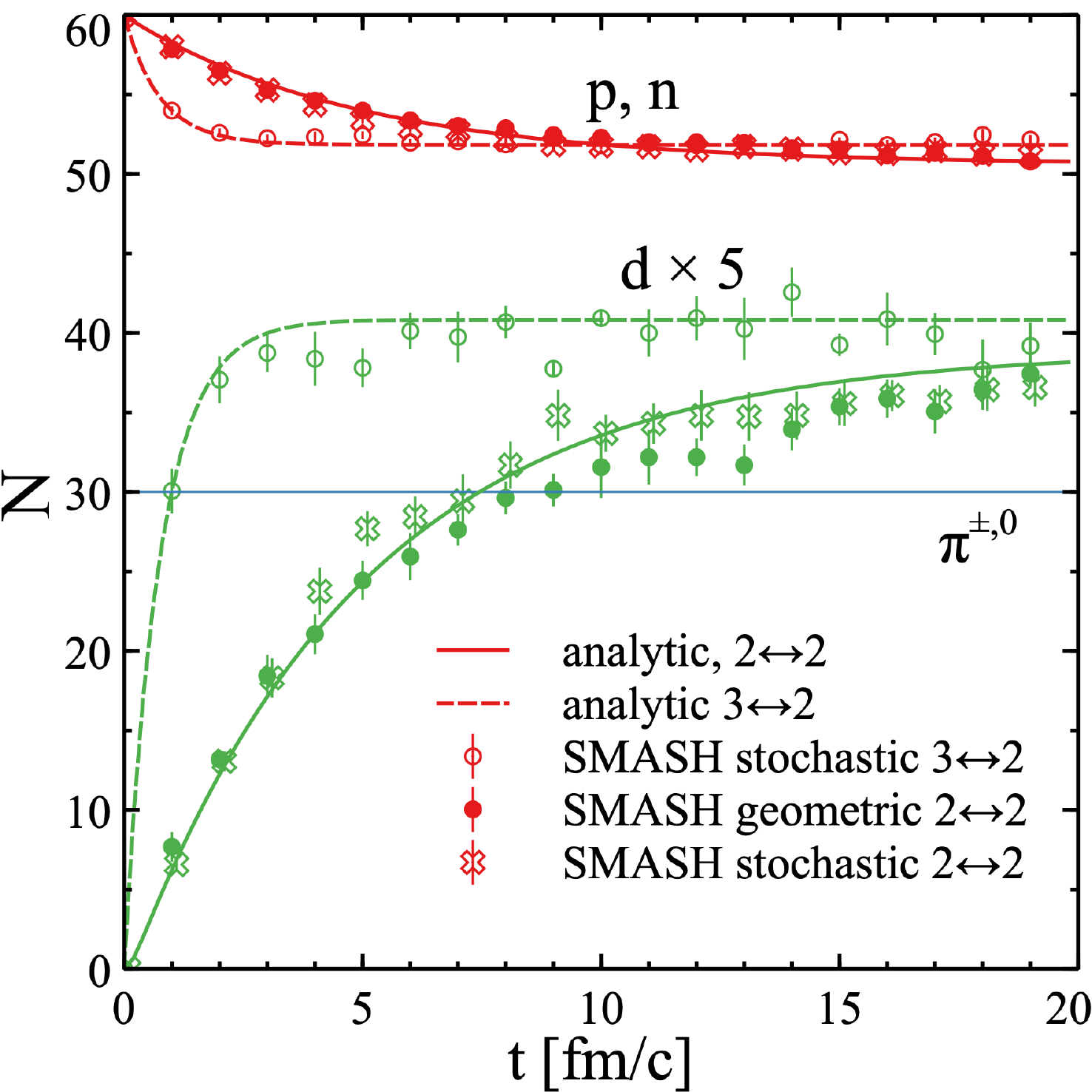}
    \caption{Evolution of particle yields in a box, analytical calculation (lines) is compared to simulations (symbols).}
    \label{fig:box_rates}
\end{figure}
The box size is set to be $V = (10$ fm $)^3$, particles are initialized uniformly in coordinate space and according to a Boltzmann distribution in momentum space with temperature $T = 0.155$ GeV. The initial multiplicities are chosen to be 30 for each pion species and 60 for each nucleon species. The cross sections of $\pi d$ and $N d$ are the ones described in \cite{Oliinychenko:2018ugs} and taken further for Au+Au simulations. The $\pi d \leftrightarrow NN$ reaction is switched off for the box test. The only allowed reactions are elastic collisions and the reactions present in Eq.~(\ref{eq:piNddprime_rates}). In Eq.~(\ref{eq:piNddprime_rates}) we assume that the temperature stays constant over time. In general, this does not have to be true in the box simulation, but we check by fitting the momentum distributions that the temperature in fact stays constant.

The comparison between the particle yields of the analytic calculation and the simulations is presented in Figure~\ref{fig:box_rates}. Since the box is only filled with $\pi$ and $N$ in the beginning, the production of $d$ over time is observed. The simulation agrees for the two different reaction treatments for $d$ production ($2\leftrightarrow 2$ with resonance and direct $2\leftrightarrow 3$) as well as employing either of the two criteria (geometric and stochastic criterion). 
The agreement allows not only to validate equilibration to the correct yields, but also the equilibration process itself. 
It is separately verified that detailed balance principle is fulfilled for all allowed reactions once the yields are equilibrated. 

Physically more interesting than the verification of the different reaction treatments, is the difference observed between the now possible 3-to-2 treatment and the modeling of the same reaction via 2-to-2 reaction. Employing direct multi-particle reactions leads to a significantly faster rise of the deuteron yield. Consequently the equilibrated yields for $d$ (and $p$) are reached significantly earlier, as predicted by the rate equations. The impact of this observed faster equilibration of the system, when employing direct 3-to-2 reactions in an expanding medium created in heavy-ion collisions, is discussed throughout the remainder of this work.

A similar result is presented for 3-to-1 reactions as part of Appendix~\ref{sec:app-31}.

\section{Deuteron production in Au+Au collisions \label{sec:results}}

In the following, the results for the deuteron production in Au+Au collisions at $\sqrt{s_{\mathrm{NN}}} = 7.7$ GeV are discussed employing the newly introduced 3-to-2 reactions. All presented results display the deuteron evolution in the afterburner stage of the hybrid approach. Special emphasis is placed on the difference to the calculations modeling the same process with a 2-to-2 reaction chain including the $d'$ resonance. Furthermore, two scenarios at particlization are distinguished in the results. In one case deuterons are assumed to be produced in the hydrodynamic stage of the collision (\emph{with d at particlization}), in the other case no deuterons are present at the start of the afterburner calculation (\emph{without d at particlization}). The distinction allows to compare two different pictures for deuteron production. The thermal model-like picture, where $d$ are produced early at high temperatures and the coalescence-like picture, where $d$ are assumed to be formed at later times.

\begin{figure}[htb]
    \centering
    \includegraphics[width=0.45\textwidth]{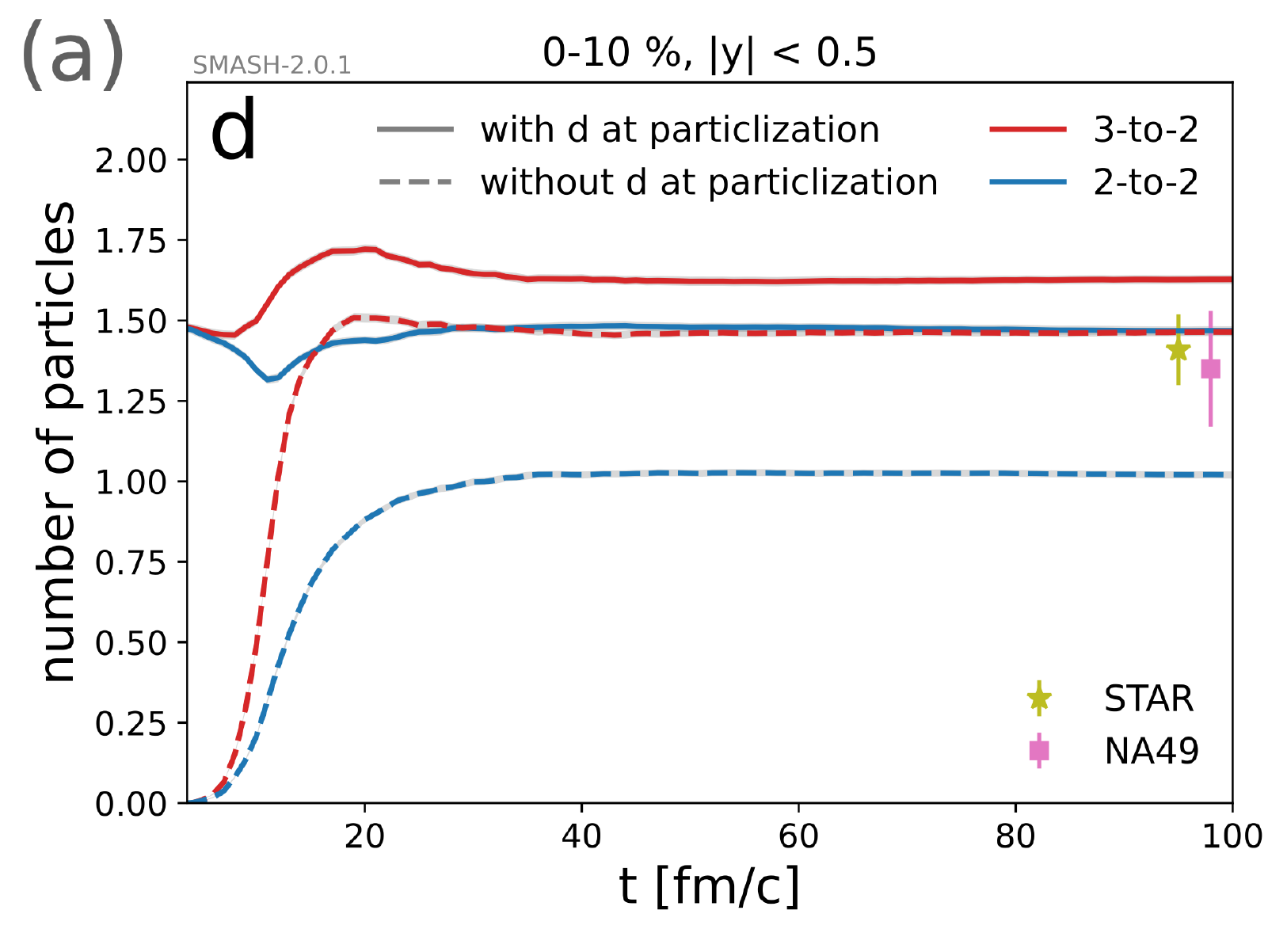}
    \includegraphics[width=0.45\textwidth]{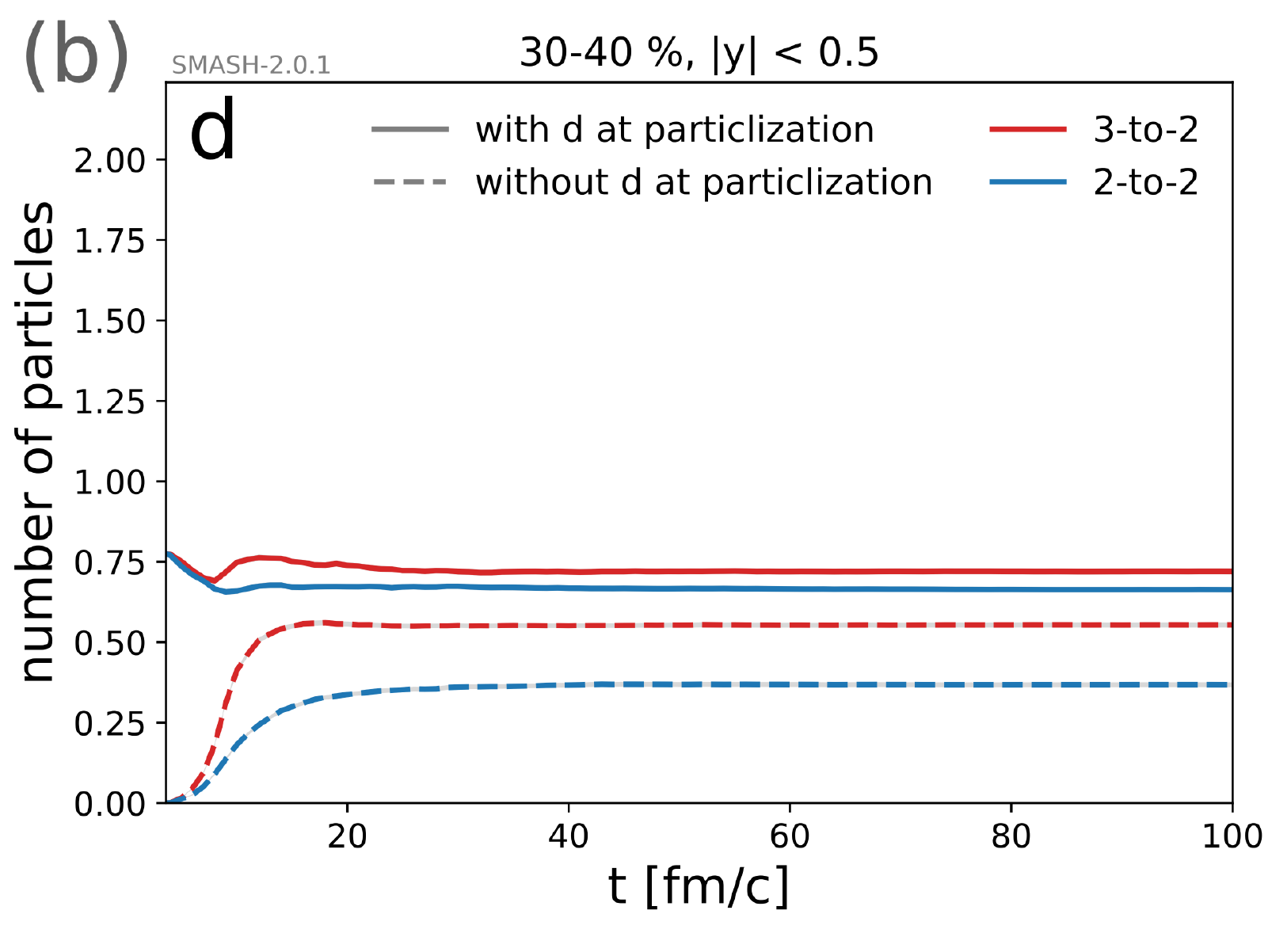}
    \caption{Evolution of mid-rapidity deuteron yields in Au+Au afterburner stage. (a): 0-10\% centrality class. (b): 30-40\% centrality class. Experimental data from \cite{Adam:2019wnb, Anticic:2016ckv}.}
    \label{fig:d-mult}
\end{figure}
\begin{figure}[h]
    \centering
    \includegraphics[width=0.45\textwidth]{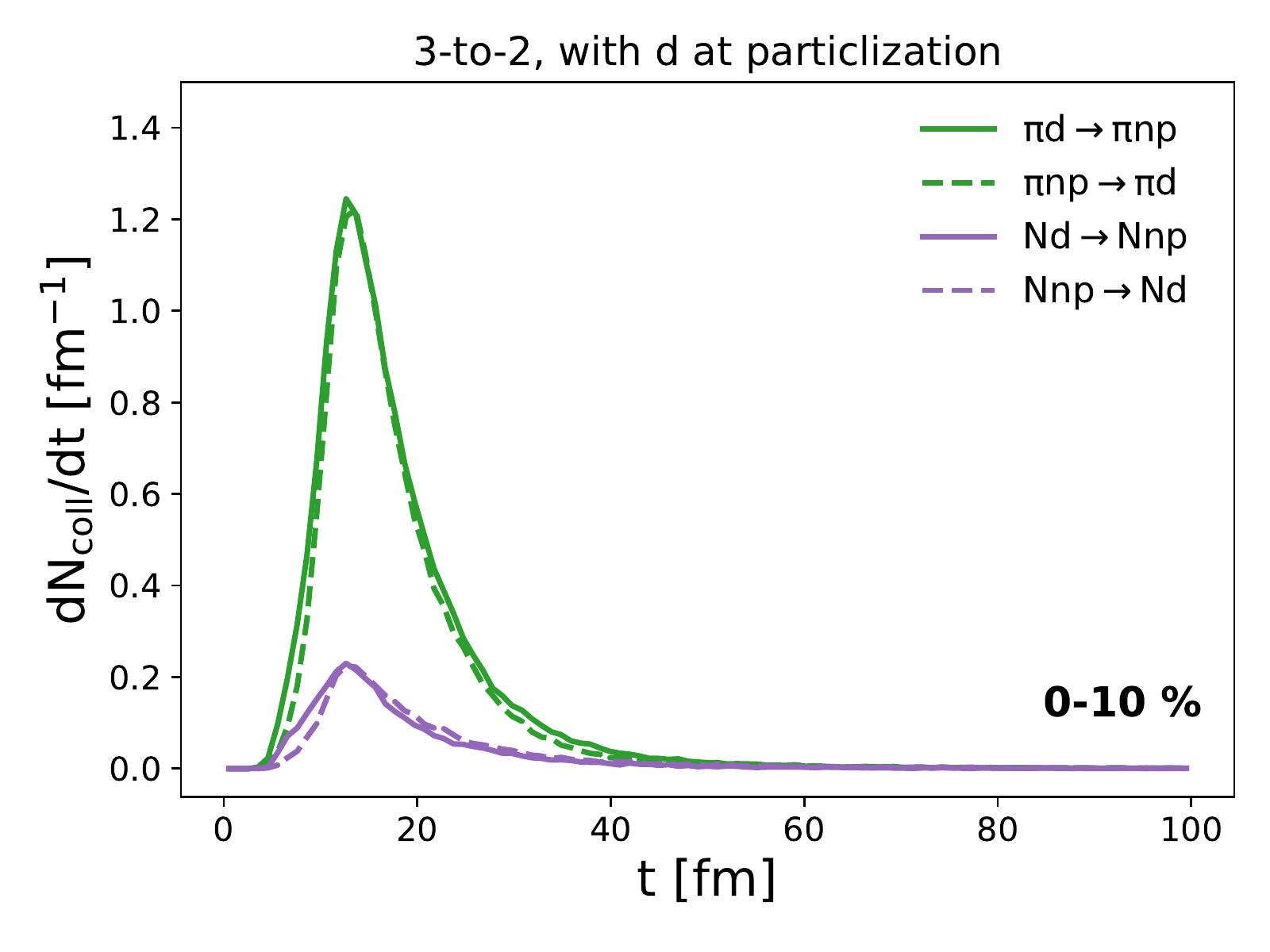}
    \includegraphics[width=0.45\textwidth]{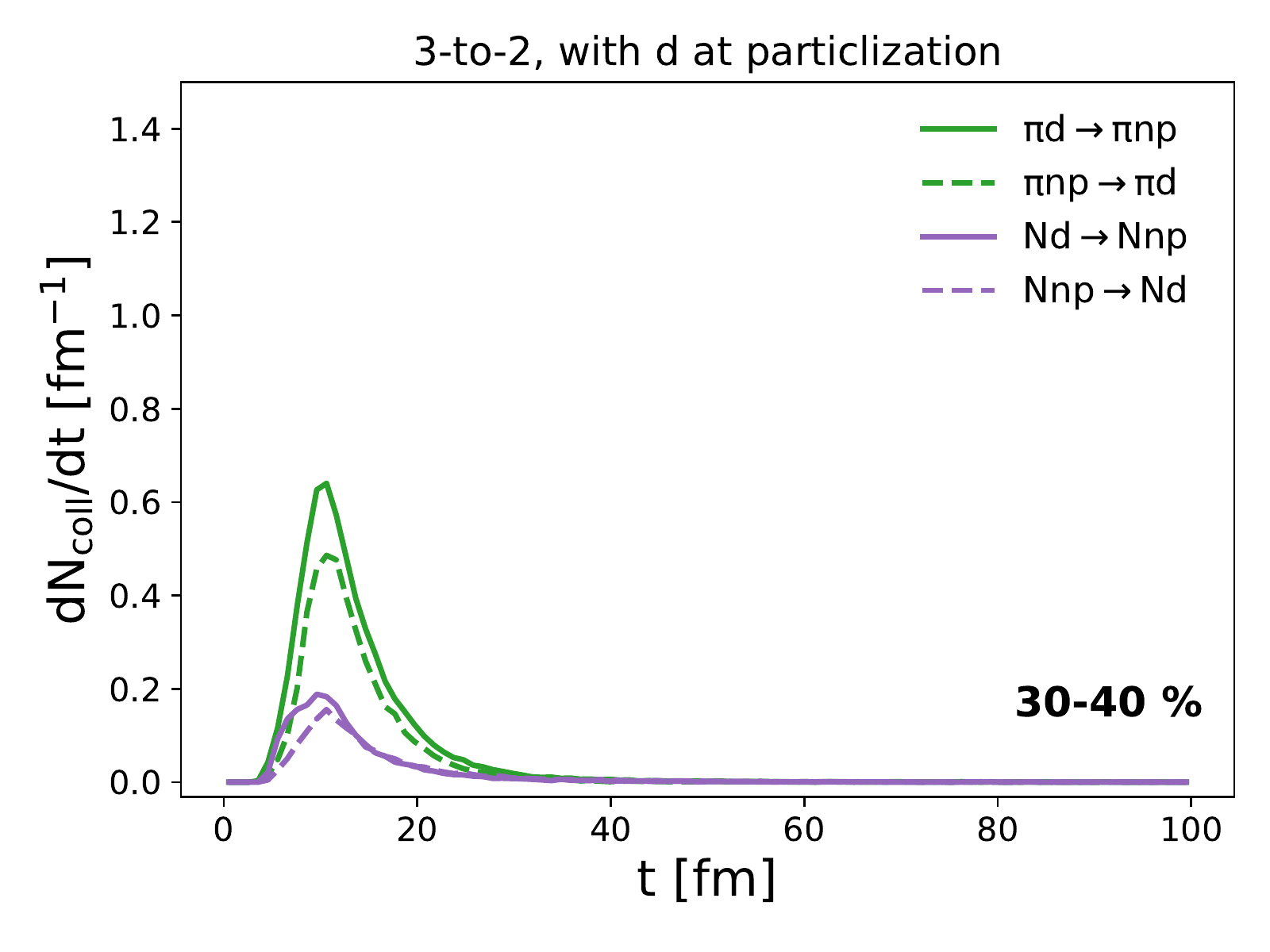}
    \caption{Evolution of $3\leftrightarrow2$ reaction rates in Au+Au afterburner with d at particlization for 0-10\% and 30-40\% centrality class.}
    \label{fig:d-rates-w}
\end{figure}
\begin{figure}[htb]
    \centering
    \includegraphics[width=0.45\textwidth]{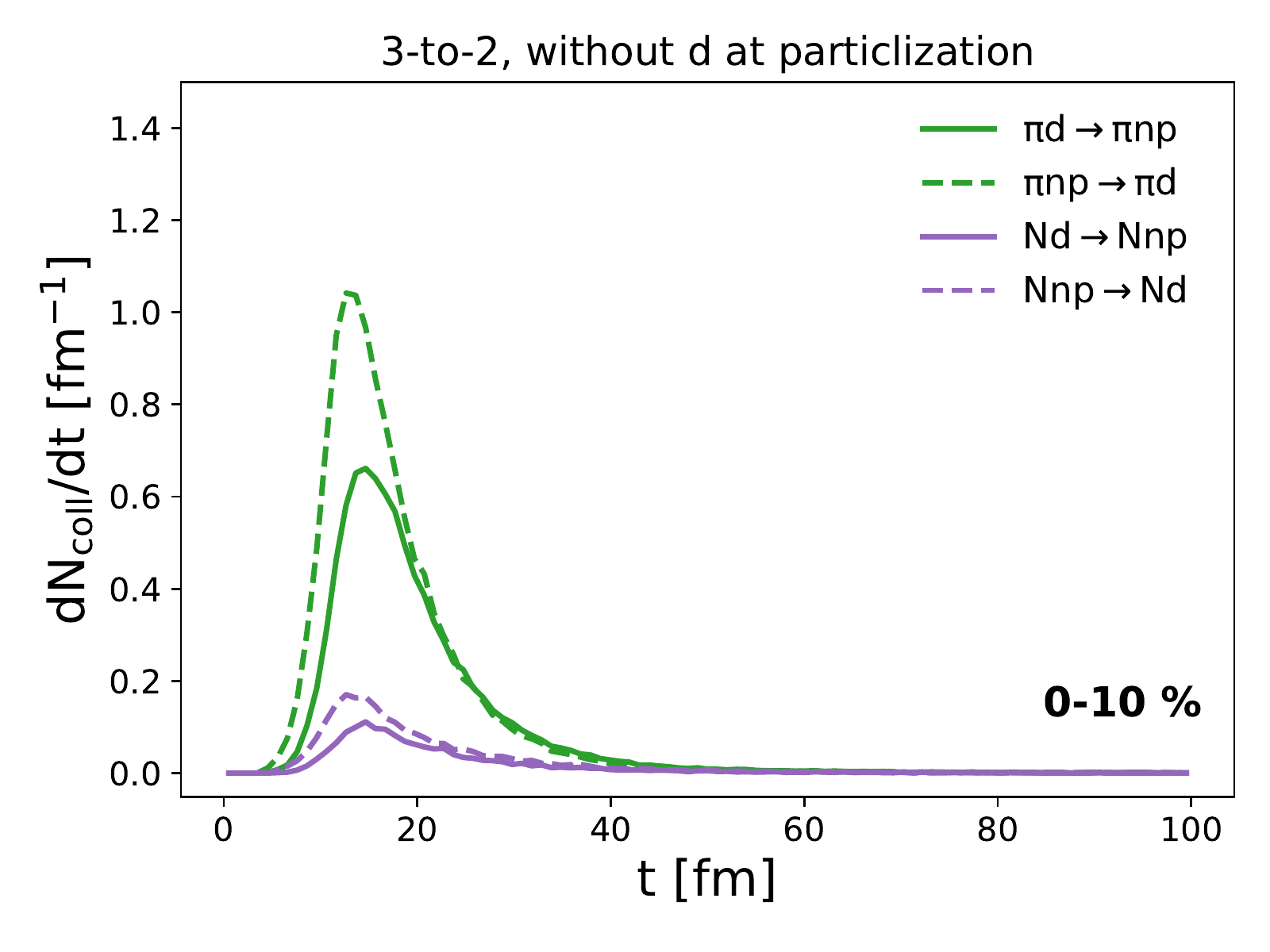}
    \includegraphics[width=0.45\textwidth]{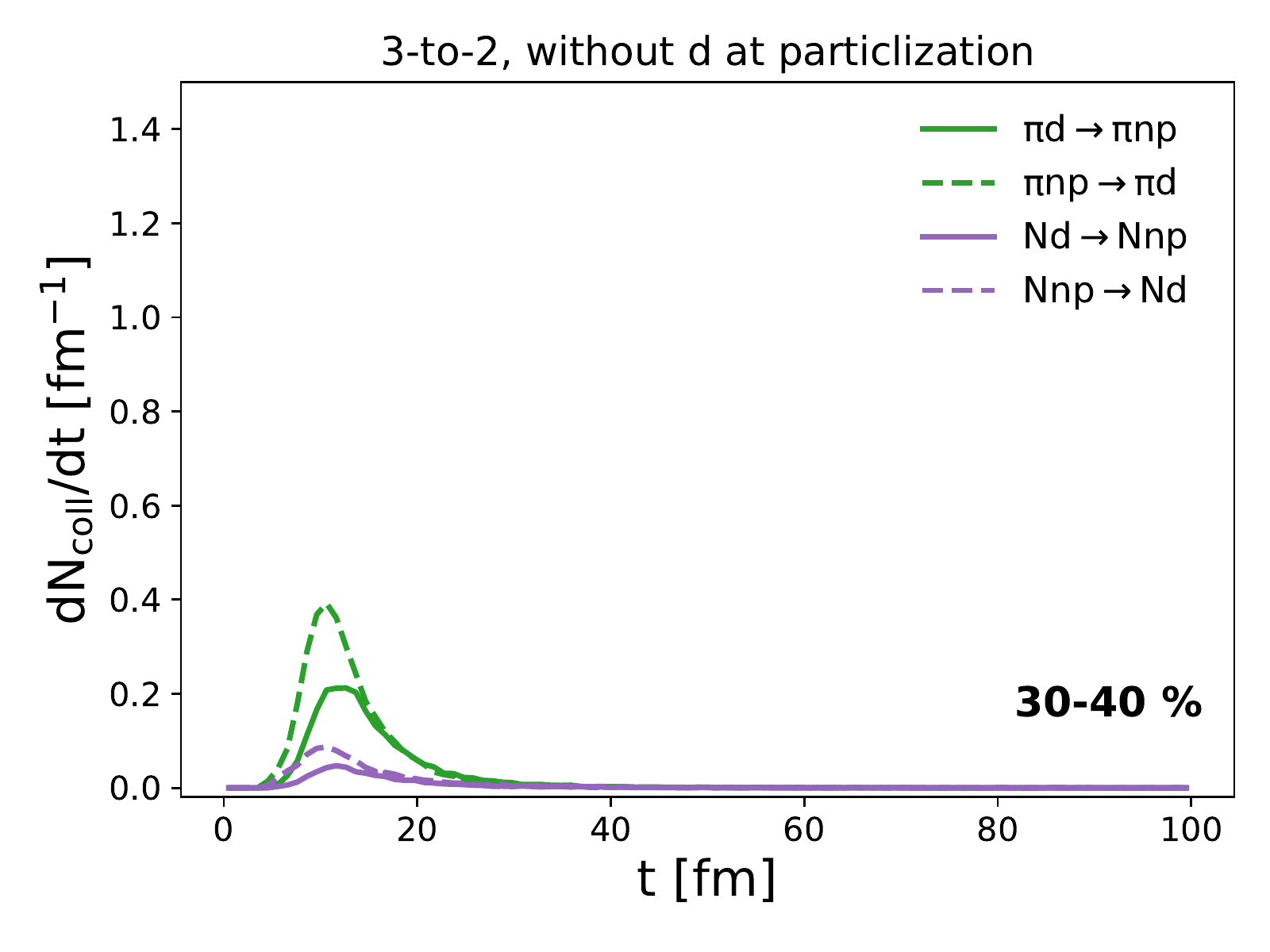}
    \caption{Evolution of $3\leftrightarrow2$ reaction rates in Au+Au afterburner without d at particlization for 0-10\% and 30-40\% centrality class.}
    \label{fig:d-rates-wo}
\end{figure}

Figure~\ref{fig:d-mult} shows the number of deuterons propagated in the afterburner stage over time. The $d$ production is enhanced when employing direct 3-to-2 reactions. Especially in the case with no $d$ at particlization, a more rapid increase of the $d$ number is observed, which drives the number close to the case with $d$ at particlization. The final number of deuterons is almost identical for the two particlization scenarios. The remaining difference is on the order of experimental errors. The difference is smaller for the calculation with the 3-to-2 reactions in comparison to the 2-to-2 approach. Those findings are understandable considering the above observed faster equilibration when employing multi-particle reactions. The 3-to-2 reactions drive the system faster to statistical equilibrium before it freezes out due to its expansion. The expansion is also the reason why the yields without $d$ at particlization are not in full agreement with $d$ at particlization, the d reactions seize too quickly due to the cooling before enough $d$ can be produced (cf. Figure~\ref{fig:d-rates-w} and \ref{fig:d-rates-wo}). Both particlization scenarios are also in agreement with the experimental values for 0-10\% centrality, which shows that $d$ yield is possible to understand in terms of multi-particle catalysis reaction being the main production mechanism.

Comparing the two presented centrality classes in Figure~\ref{fig:d-mult}, the more peripheral collisions studied with the 30-40\% class naturally produce less $d$ in general. The two deuteron reaction treatments also produce more similar yields in the case with $d$ at particlization.

\begin{table}
\centering
\begin{tabular}{ c | c }
 Centrality &  $\rm N_{d,\,equil}^{without}  \,\big/\,N_{d,\,equil}^{with}$ \\ [0.7ex]
 \hline
 0-10\% & 0.900 \\
 10-20\% & 0.860 \\
 20-30\% & 0.795 \\
 30-40\% & 0.769 \\
 40-50\% & 0.731 \\
 50-60\% & 0.759 
\end{tabular}
\caption{Ratio of the equilibrated $d$ yields for the two particlization scenarios.}
\label{tab:ratio-w-vs-wo-d}
\end{table}

Going from central to more peripheral collisions, the equilibrated yield  without $d$ at particlization ($\rm N_{d,\,equil}^{without}$) in comparison to the final yield with $d$ at particlization ($\rm N_{d,\,equil}^{with}$) is less in agreement as Table~\ref{tab:ratio-w-vs-wo-d} shows. The smaller medium created in peripheral collision seems to suppress the full statistical equilibration of the system before freeze-out when all $d$ are produced in the late (afterburner) stages.

Employing the stochastic criterion for the 2-to-2 reaction chain (not shown), yields the same results as obtained with the geometric criterion. The number of test particles has been chosen to be $N_{\rm test} = 20$ for the results. It is separately verified that higher numbers of test particles are matching the displayed results.

The evolution of the $d$ yield is the result of the competing $3\rightarrow2$ formation and $2\rightarrow3$ break-up reaction rates shown in Figure~\ref{fig:d-rates-w} for the case with $d$ at particlization and in Figure~\ref{fig:d-rates-wo} for without $d$ at particlization. Forward and backward (direct) 3-to-2 reactions rates are close for central collisions in Figure~\ref{fig:d-rates-w}, but as also seen for the yield some time is necessary before they are close to being equilibrated. The collision rates also clearly indicate the dominance of the $\pi$ catalysis reactions. This underlines the necessity to include the $\pi$ in addition to the $N$ reactions for this beam energy, which is the main extension compared to \cite{Danielewicz:1991dh}.
The $\pi$ reactions are also closer to being equilibrated than the $N$ reactions. For more peripheral reactions and with this a smaller medium, the $2\rightarrow3$ rate is dominating over the formation reactions. This relatively lower 3-to-2 reaction rate again hints at incomplete statistical equilibration in the smaller system.
Considering the calculation without $d$ in Figure~\ref{fig:d-rates-wo}, the $3\rightarrow2$ reaction dominates for all rates, as expected by the rapid rise of $d$ numbers at the beginning of the evolution. 
The reaction rate figures also allow to pinpoint a chemical freeze-out at least for the shown $3\leftrightarrow2$ d reactions at around $50$ fm for central collisions (0-10\%). Since the system is smaller for more peripheral collisions (30-40\%), the freeze-out is earlier at around $30$ fm. Note that while the $NN\leftrightarrow\pi d$ reaction is also included in the calculation, its contribution is only sub-leading, even compared to the $N$ catalysis reaction, and therefore the rate is not shown.

\begin{figure}[htb]
    \centering
    \includegraphics[width=0.45\textwidth]{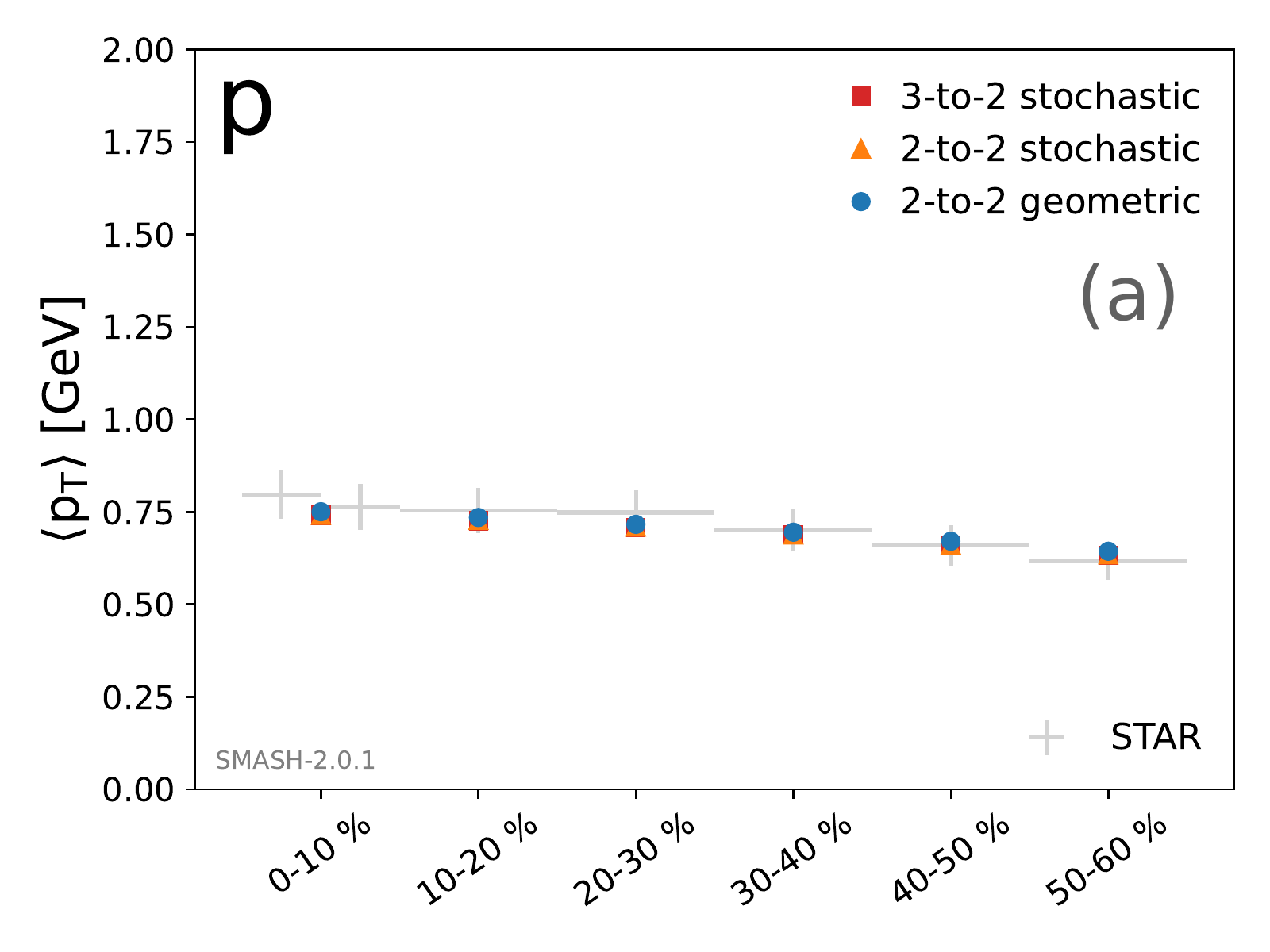}
    \includegraphics[width=0.45\textwidth]{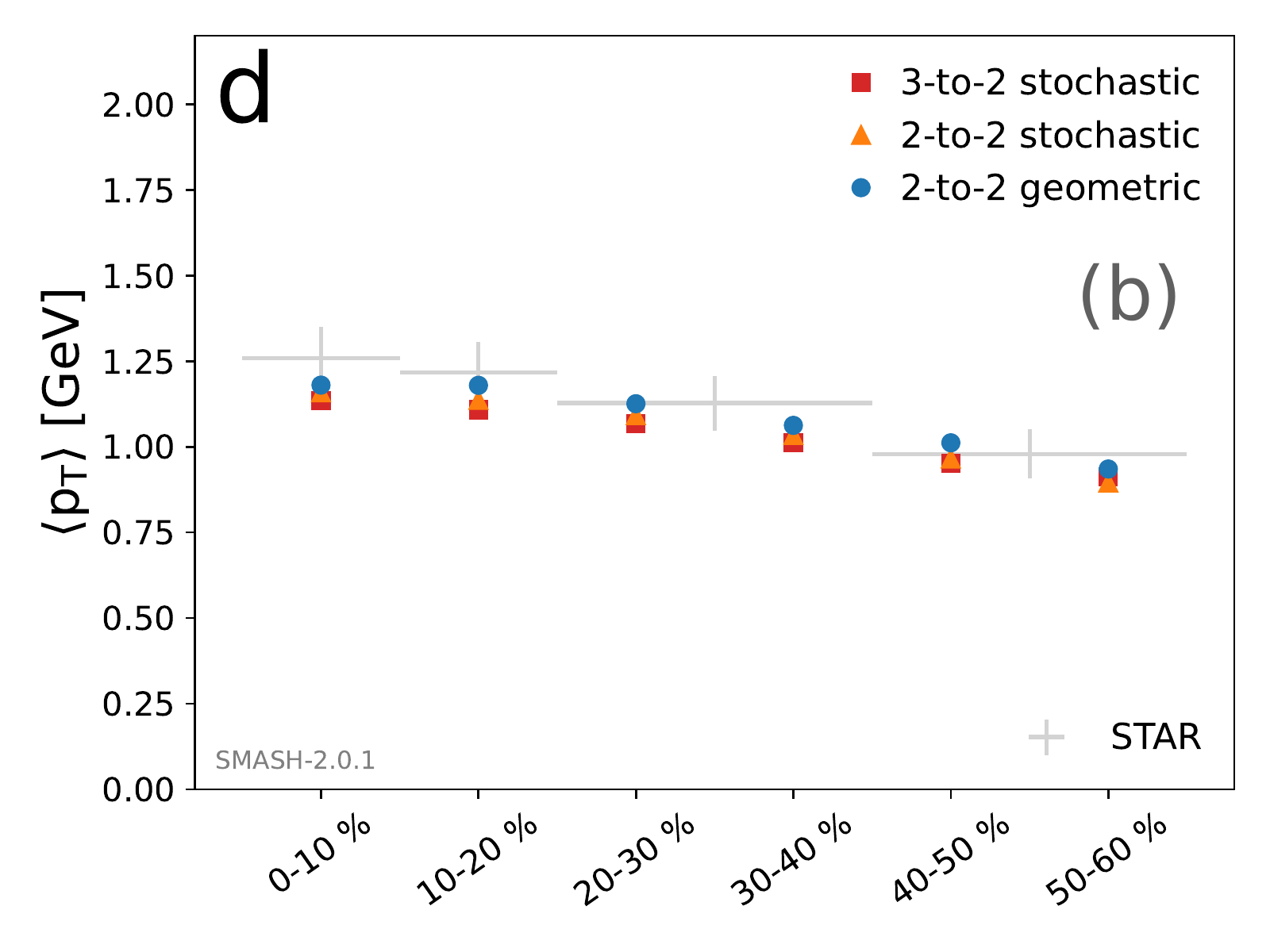}
    \caption{Average transverse momentum for protons ($\rm p$) and deuterons ($\rm d$). Experimental data from \cite{Adamczyk:2017iwn, Adamczyk:2016gfs}.}
    \label{fig:mean-pt}
\end{figure}

In addition to the yields, first the average transverse momentum is presented for protons and deuterons in Figure~\ref{fig:mean-pt} for 6 centrality classes. The mean-$p_T$ slightly declines towards more peripheral collisions. Here and in the following, three treatments for the deuteron catalysis reactions are compared: calculating with the 2-to-2 reaction chain for the geometric (blue round points) and the stochastic (orange triangles) criterion as well as direct 3-to-2 reactions (red squares) that are only possible with the stochastic criterion. The mean-$p_T$ results are unaffected by the different approaches and all agree with the available experimental data~\cite{Adam:2019wnb, Anticic:2016ckv} within errors for both, $p$ and $d$, validating the transverse dynamics of the calculations. Note that this confirms the previous findings in~\cite{Oliinychenko:2020znl}, where the here employed hybrid approach was carefully constrained by an extensive experimental data set.

\begin{figure}[htb]
    \centering
    \includegraphics[width=0.45\textwidth]{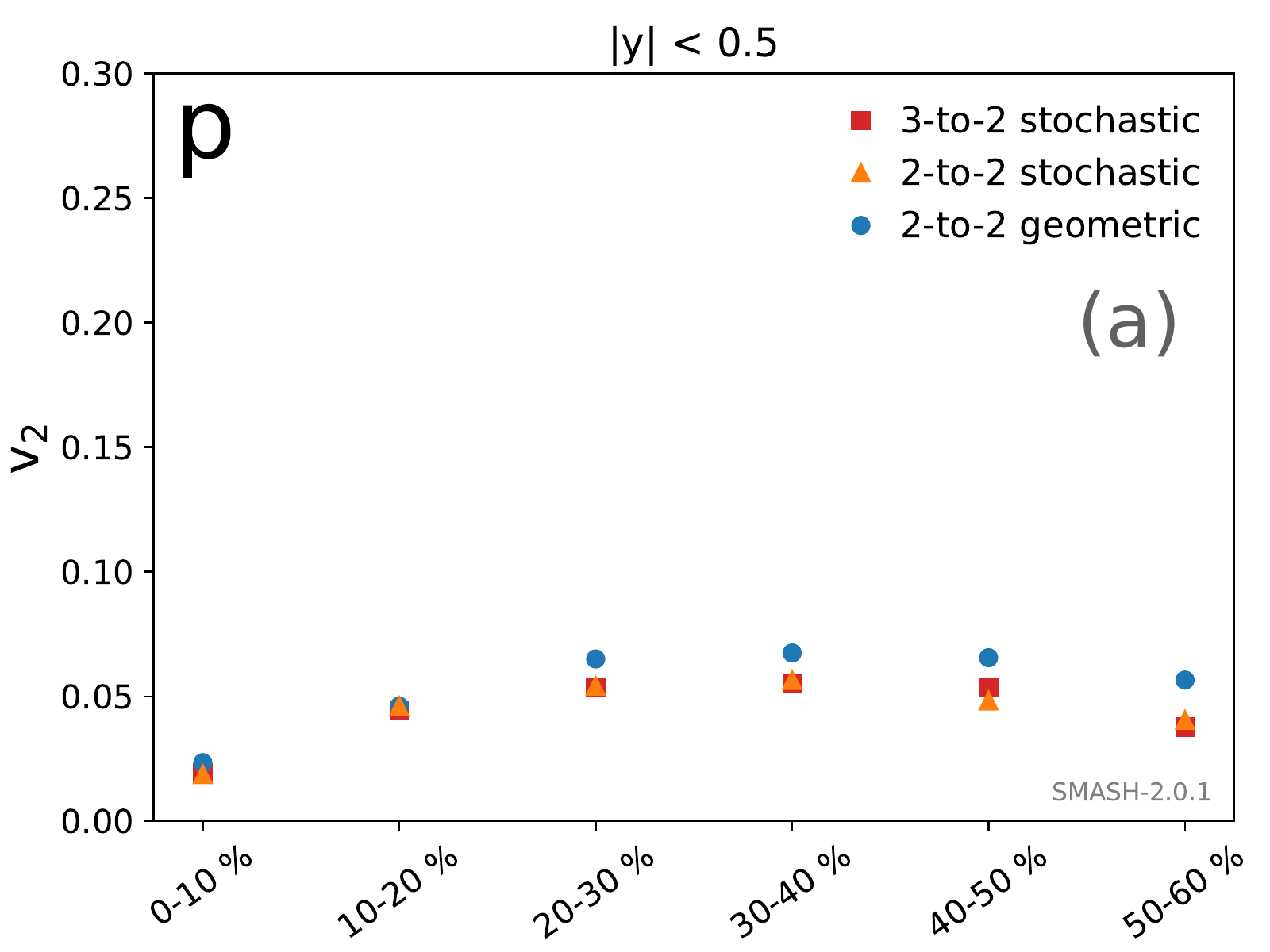} \\
    \includegraphics[width=0.45\textwidth]{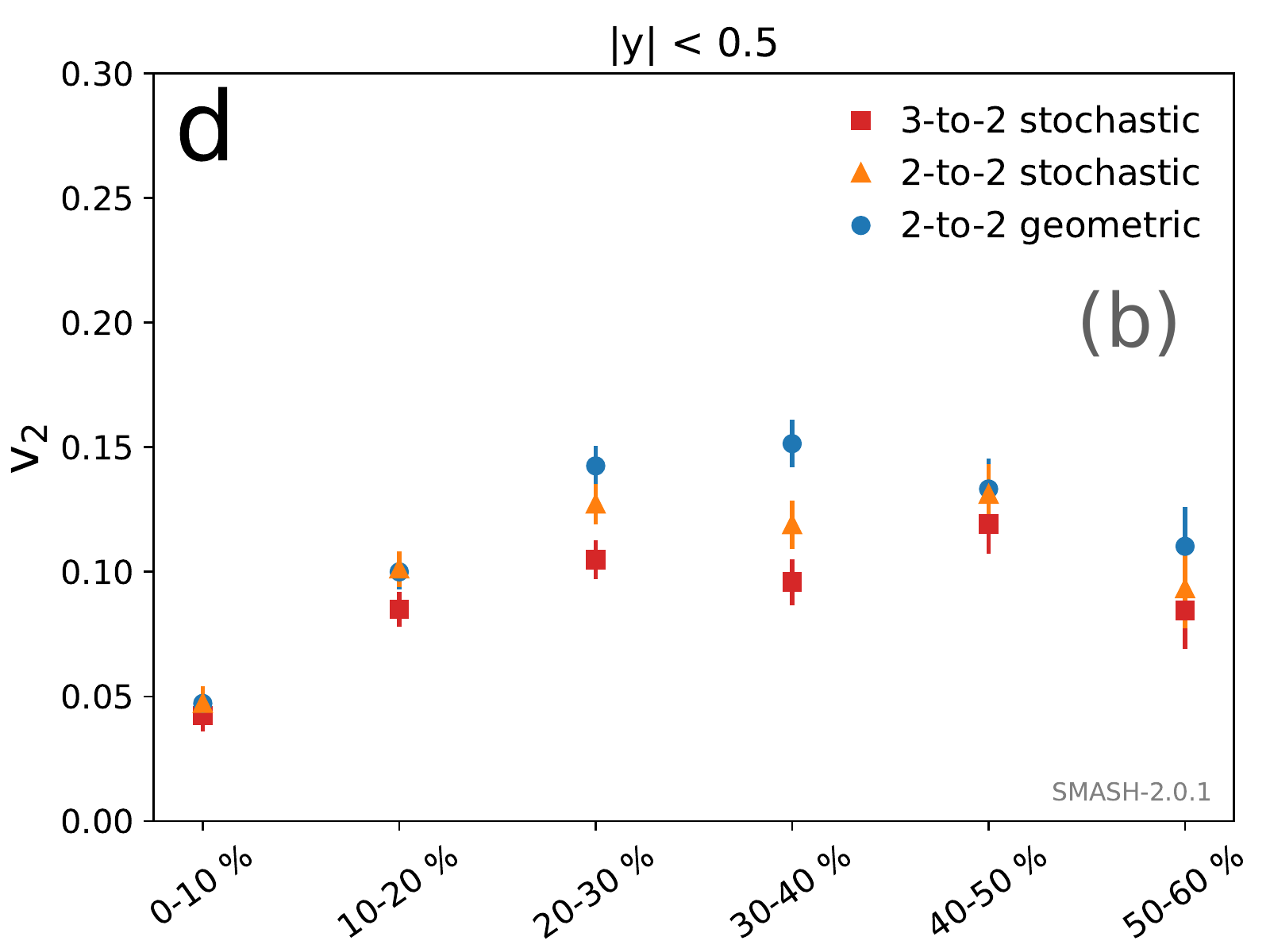}
    \caption{Integrated elliptic flow for protons ($\rm p$) and deuterons ($\rm d$).}
    \label{fig:int-v2}
\end{figure}
Figure~\ref{fig:int-v2} presents the results for the elliptic flow ($v_2$) for $p$ and $d$ for the same 6 centrality classes, as a more sensitive probe of the momentum distribution. Due to limited statistics of the calculation only the integrated $v_2$ is presented, which still allows to contrast the different approach for the $d$ reactions. Even though no experimental data is available for this observable, the order of magnitude of around $0.1$ is comparable to the $p_T$ dependent $v_2$ reported in \cite{Adamczyk:2016gfs}. Regarding the integrated $v_2$ of $d$ in the bottom panel of the figure, the different reaction treatments are found to have an effect. While for central collision they agree, for more peripheral collisions the elliptic flow is decreased for the 3-to-2 reactions and by employing the stochastic criterion. The latter is also found for the protons in the upper panel. The additional effect for 2-to-2 reactions for the two different collision criteria might hint at limitations of the stochastic criterion for the small systems in peripheral collisions. However, a more uniform, thermalized medium and a subsequent lower $v_2$ might also be expected when employing multi-particle reactions.
\begin{figure}[htb]
    \centering
    \includegraphics[width=0.45\textwidth]{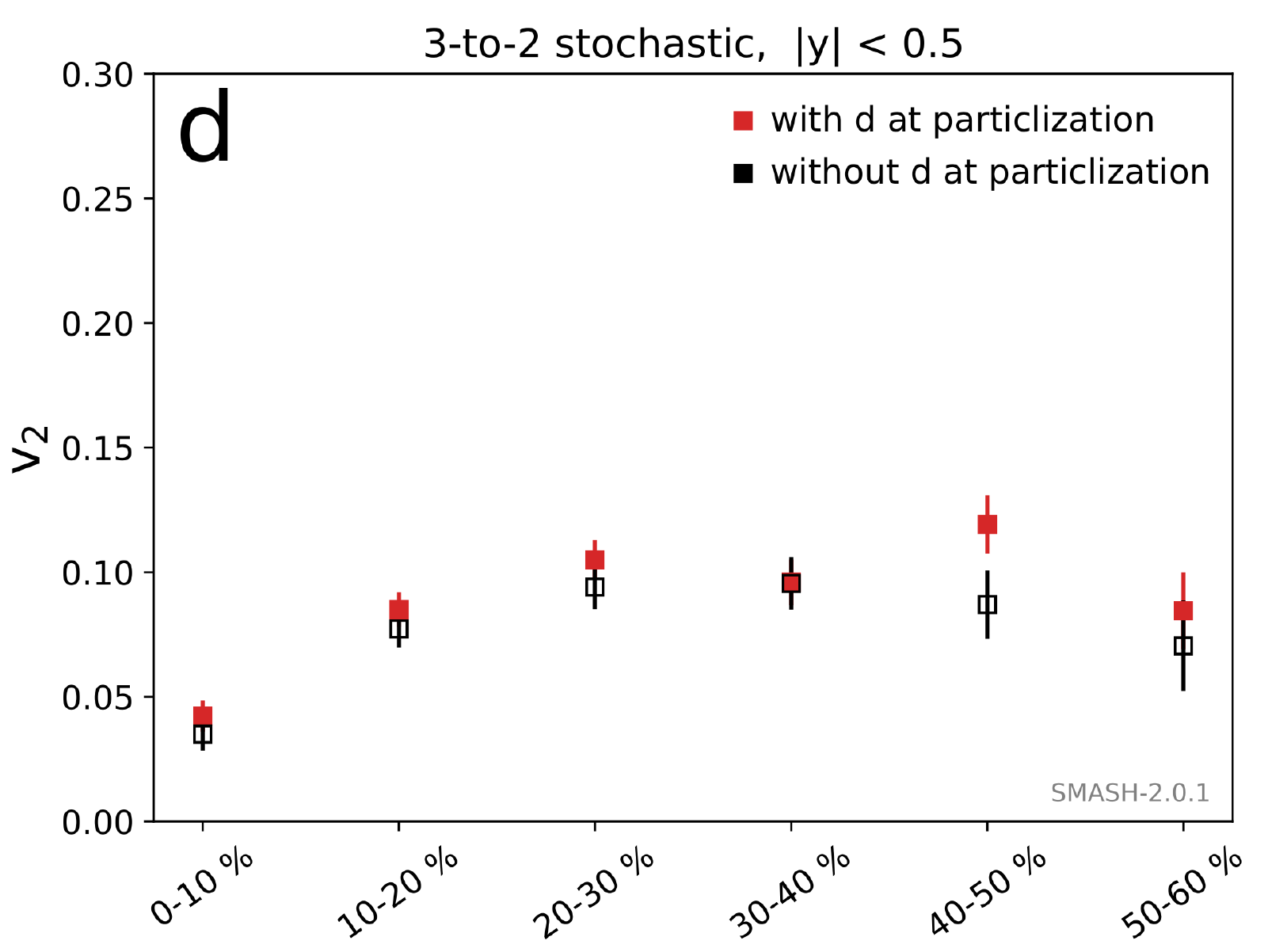}
    \caption{Integrated elliptic flow for deuterons with and without d at particlization.}
    \label{fig:int-v2-comp-d-part}
\end{figure}
In Figure~\ref{fig:int-v2-comp-d-part} the impact of $d$ solely being produced in the afterburner on the elliptic flow is studied. Observing a clear difference here would potentially enable to disentangle the two pictures for $d$ production times. Even though a clear difference is not found, a small decrease is seen for the case without $d$ at particlization. The difference is not significant for all  centrality classes, the effect for the average over all centralities, however, is significant.

\section{Conclusion} \label{sec:conclusions}

This work presents the first application of multi-particle reactions in the hadronic transport approach \texttt{SMASH} by employing a stochastic collision criterion. This is of major importance since it allows to treat multi-particle reactions while fulfilling detailed balance.

A comprehensive derivation of the stochastic collision criterion is shown for different reaction classes, from an one- and two-body initial state to the multi-particle reactions with 3 particles. The derivation for 3-body reactions focuses on the inverse reaction for meson Dalitz decays ($3\rightarrow 1$) and the deuteron catalysis reactions ($3\leftrightarrow 2$): $\pi p n \leftrightarrow \pi d$ and $N p n \leftrightarrow N d$. The stochastic collision criterion allows to avoid the introduction of an artificial resonance by treating the deuteron $3\rightarrow 2$ catalysis reactions in one-step while simultaneously fulfilling detailed balance. 

The newly-introduced stochastic criterion is validated by an agreement with the analytic expectation for the two-body collision rate in a box setup with elastic collisions. In addition, the particle multiplicities for heavy-ion reactions at different beam energies are reproduced, when including the same reactions as in the default \texttt{SMASH} version. An analytic expression for the chemical equilibration is derived with rate equations. The stochastic multi-particle reactions for the deuteron catalysis are validated by agreeing with the analytic results, exhibiting the correct chemical equilibration and detailed balance. As predicted by the rate equations, the time required to equilibrate is significantly reduced for the direct $3\leftrightarrow 2$ treatment compared to the previous $2\leftrightarrow 2$ multi-step process. The finding of a faster equilibration employing multi-particle reactions is also confirmed in 3-to-1 reactions (Figure~\ref{fig:3to1-box}).

Studying the deuteron production in gold-gold collision at a beam energy of 7.7 GeV in a hybrid approach, the faster equilibration process of the multi-particle reactions leads to a more rapid increase in the deuteron yield before the system freezes out chemically due to the expansion of the system. The $d$ yield is consequently enhanced. 
The difference in the final number of $d$, when comparing the scenarios of $d$ being produced in the hydrodynamic stage or just in the hadronic afterburner, is greatly reduced when employing multi-particle reactions due to three-body reaction driving the $d$ faster to statistical equilibrium.
The yield agrees with the experimental data if the nuclei are produced at time of particilization or not. Hereby confirming the previous findings when employing the slower equilibrating two-body reaction chain involving the fake $d'$ resonance~\cite{Oliinychenko:2018ugs,Oliinychenko:2020znl}. 
In addition, a decrease in the elliptic flow is found when employing the stochastic criterion as well as multi-particle reactions for more peripheral collisions. Similarly a small decline in flow for all centralities is reported, if all $d$ are produced during the late (afterburner) stages of the collision. No dependency is found for mean transverse momentum, where an agreement with experimental data for all centrality classes is seen.

The stochastic criterion for multi-particle reactions also opens the possibility to investigate other reactions in the future. Of particular interest is the theoretically straightforward extension of this work to include other light nuclei like (hyper-) triton and $\rm{^4He}$. While the conclusions employing the $2\leftrightarrow 2$ reaction treatment involving the artificial $d'$ are confirmed by this work, introducing additional fake resonances like $t'$ is tedious and would certainly enhance the found differences to the theoretical preferable direct $3\leftrightarrow 2$ treatment. In addition, the stochastic reaction approach is directly extendable to study larger $n$-particle reactions ($n>3$). Here, the study of the back-reaction of $p\overline{p}$ annihilation ($5\pi\rightarrow p\overline{p}$) will be especially interesting.

\begin{acknowledgments}
The authors thank Chun Shen for providing the hydrodynamic particlization hypersurfaces. J.S. and J.M.T.-R. were funded by the Deutsche Forschungsgemeinschaft (DFG, German Research Foundation) through project number 315477589 – TRR 211 (Strong-interaction matter under extreme conditions). D.O.  was  supported  by the  U.S.  DOE  under  Grant  No. DE-FG02-00ER4113. J.M.T.-R. furthermore acknowledges support from the Deutsche Forschungsgemeinschaft (DFG, German research Foundation) through project no.  411563442 (Hot Heavy Mesons).  Computational resources were provided by Goethe-HLR cluster. 
\end{acknowledgments}

\appendix

\section{Reaction probability for a $n \rightarrow m$ process \label{sec:app-pnm}}

The stochastic collision criterion uses a reaction probability which depends on the number and type (stable particle or resonances) of incoming and outgoing states. For a generic reaction from $n$ to $m$ particles (we will denote with primed indices the states in the final state) the reaction probability reads
\be \label{eq:Pnm}  
P_{n \rightarrow m} = \frac{\Delta N_{\textrm{reactions}}}{\displaystyle\prod_{j=1}^n \Delta N_j} 
\ . \ee

Using the Boltzmann equation to quantify the number of reactions in a phase-space element during a time $\Delta t$ we finally obtain,
\be \label{eq:Pnmbig}
P_{n\rightarrow m} = \frac{1}{{\cal S}'!} \frac{\Delta t}{(\Delta^3 x)^{n-1}} \frac{1}{\displaystyle\prod_{j=1}^n 2E_j} \int d\Phi_m \ \overline{|T_{n\rightarrow m}|^2} \ , 
\ee
where ${\cal S}'$ is the number of identical particles in the final state (to avoid double counting when integrating the phase-space variables), e.g. if the final state contains $N p n$, then ${\cal S}'!=2$. $\overline{|T_{n\rightarrow m}|^2}$ is the scattering matrix squared of the process, summed over final internal states and averaged over the initial ones,
\be \overline{|T_{n\rightarrow m}|^2} = \frac{1}{\displaystyle\prod_{j=1}^n g_j} \ \sum_{\substack{\textrm{initial}\\ \textrm{final}}}  |T_{n\rightarrow m}|^2 \ , \ee
where $g_j$ is the internal degeneracy of the state $j$. Since we distinguish isospin states, it will be used to account for spin degeneracy $g_j=2\mathfrak{s}_j+1$, where $\mathfrak{s}_j$ is the spin of the state.

In Eq.~(\ref{eq:Pnmbig}) the factor $d\Phi_m$ is the $m-$particle phase-space element,
\be \label{eq:phasespace}
d\Phi_m =  (2\pi)^4\delta^{(4)} \left(P-\sum_{k=1}^m p_k \right) \prod_{k=1}^m d\Gamma_k \ , 
\ee
where $P$ is the total 4-momentum of the reaction. For stable particles we will use (keeping implicit the particle subindex $k$ for simplicity)
\be d\Gamma = \frac{d^3p}{(2\pi)^3 2E} \ , \ee
with $E^2=p^2+m^2$, while for resonances
\be d\Gamma = \frac{d^4p}{(2\pi)^4} \frac{\pi}{\sqrt{s}} {\cal A} (\sqrt{s})=\frac{dM}{2} \frac{1}{E_{M}}  \frac{d^3p}{(2\pi)^3}  {\cal A}(M) \ , \label{eq:spectfunc} \ee
where $M=\sqrt{s}=\sqrt{ p^{0,2}-p^2}$, $E_{M}^2=p^2+M^2$ and ${\cal A} (M)$ is the spectral function of the state, normalized to
\be \int_0^\infty dM {\cal A} (M) = 1 \ . \ee
In the limit of very narrow resonance one recovers the stable particle case by doing ${\cal A}(M)=2M\delta(M^2-m^2)$.

Usually one assumes for simplicity that the scattering amplitude (when it is unknown) only depends on the initial center-of-mass energy. In such cases one can pull it out of the integral over the final phase-space.
When this is used, the integrated $m-$body phase space reads for $m=1,2,3$,
\begin{align} 
  \Phi_1 (M^2) & = \frac{\pi}{M} {\cal A}(M) \ , \label{eq:Phi1} \\
  \Phi_2 (M^2;m^2_1,m^2_2) & = \frac{\lambda^{1/2}(M^2,m^2_1,m^2_2)}{8\pi M^2} \ , \label{eq:Phi2} \\
  \Phi_3 (M^2;m^2_1,m^2_2,m^2_3) & = \frac{1}{2\pi} \int_{(m_1+m_2)^2}^{(M-m_3)^2} ds_1 \Phi_2(M^2;s_1,m^2_3) \nonumber \\
  &\times \Phi_2(s_1;m^2_1,m^2_2) \ , \label{eq:Phi3}
\end{align}
where $\lambda(M^2;m^2_1,m^2_2)=(M^2-m_1^2-m_2^2)^2-4m^2_1m_2^2$ is the K\"all\'en function.

For special cases $n,m=1,2$ one can write the probabilities in terms of the decay width of interaction cross sections. Let us consider the particular cases considered in this work.

\begin{itemize}
 \item $n=1,m=2$ In this case we can introduce the decay width of a resonance (with the degeneracy and symmetry factors incorporated)
 \be 
 \Gamma_{1\rightarrow 2} (M) = \frac{1}{{\cal S}'!} \int d\Phi_2 \overline{|T_{1\rightarrow 2}|^2} 
 \ee
 to arrive to
 \be 
 P_{1\rightarrow 2} = \Delta t  \frac{M}{E_M} \Gamma_{1\rightarrow 2} (M)  \ . \label{eq:P12} 
 \ee
 \item $n=2,m=1$
 The inverse reaction reads
 \be
 P_{2\rightarrow 1} = \frac{g_{1'}}{g_1 g_2}  \frac{\Delta t}{\Delta^3 x} \frac{1}{2E_1 2E_2} \int d\Phi_1 \overline{|T_{1\rightarrow 2}|^2} \ . 
 \ee
For practical purposes we introduce the relative velocity
\be 
v_{\textrm{rel}} \equiv \frac{\lambda^{1/2}(s;m_1^2,m_2^2)}{2E_1 E_2} \ , \label{eq:vrel}
\ee
and the $2\rightarrow 1$ cross section (cf. Eq.~(39) in \cite{Weil:2016zrk})
\begin{align}
\sigma_{2\rightarrow 1} (\sqrt{s}) & \equiv \frac{g_{1'}}{g_1 g_2} \frac{8\pi^2 s}{\lambda(s,m^2_1,m^2_2)} {\cal A}(\sqrt{s}) \Gamma_{1\rightarrow 2}(\sqrt{s}) \ . \nn \\
\label{eq:xsec1}
\end{align}
The $P_{2\rightarrow 1}$ reads finally
\be P_{2\rightarrow 1} = \frac{\Delta t}{\Delta^3 x} v_{\textrm{rel}} \sigma_{2\rightarrow 1} (\sqrt{s}) \ . \ee

 \item $n=1,m=3$
The decay to three particles is similar,
\be
P_{1\rightarrow 3} = \Delta t \frac{M}{E_M} \Gamma_{1\rightarrow 3} (M) \ , 
\ee
where we introduced the decay width
\be 
\Gamma_{1\rightarrow 3} (M) = \frac{1}{{\cal S}'!} \int d\Phi_3 \overline{|T_{1 \rightarrow 3}|^2} \ . \label{eq:Gamma13}
\ee
\item $n=3,m=1$
This case can be written as
\begin{align}
P_{3\rightarrow 1} & = \left( \frac{g_{1'}}{g_1 g_2 g_3} \right) {\cal S}! \frac{\Delta t}{(\Delta^3 x)^2} \frac{\pi}{4E_1E_2E_3} \nn \\
& \times \frac{\Gamma_{1\rightarrow 3} (M)}{\
\Phi_3(M^2)} {\cal A} (M)  \ , \label{eq:P31}
\end{align}
where the factor ${\cal S}$ counts the number of identical particles in the initial state of the $3\rightarrow 1$ reaction. It appears here because the same factor is included in the definition of $\Gamma_{1\rightarrow 3}$ [called ${\cal S}'$ in Eq.~(\ref{eq:Gamma13})], and needs to be canceled.
\item $n=m=2$ This case represents a binary collision
\be P_{2\rightarrow 2} =\frac{\Delta t}{\Delta^3 x} v_{\textrm{rel}} \sigma_{2\rightarrow 2} (\sqrt{s}) \ , 
\ee
where
\be \sigma_{2\rightarrow 2} (\sqrt{s}) =  \frac{1}{{\cal S}'!} \int d\Phi_2 \frac{ \overline{|T_{2\rightarrow 2}|^2}}{2\lambda^{1/2}(s;m_1^2,m_2^2)}
\ee
\item $n=2,m=3$ The next two cases are needed for the deuteron formation/annihilation. 
\be \label{eq:P23}
P_{2\rightarrow 3} =\frac{\Delta t}{\Delta^3 x} v_{\textrm{rel}} \sigma_{2\rightarrow 3} (\sqrt{s}) \ , 
\ee
where the $2\rightarrow 3$ cross section reads
\be \label{eq:P23bis}
\sigma_{2\rightarrow 3} (\sqrt{s}) = \frac{1}{{\cal S}'!} \int d\Phi_3 \frac{ \overline{|T_{2\rightarrow 3}|^2}}{2\lambda^{1/2}(s;m_1^2,m_2^2)} \ . 
\ee
\item $n=3,m=2$
\begin{align} 
P_{3\rightarrow 2} &  =\left( \frac{g_{1'} g_{2'}}{g_1 g_2 g_3} \right) \frac{{\cal S}!}{{\cal S}'!}  \frac{\Delta t}{(\Delta^3 x)^2} \frac{E_{1'}E_{2'}}{2E_1E_2E_3} \frac{\Phi_2(s)}{\Phi_3(s)} \nonumber \\ 
&\times v_{\textrm{rel}} \sigma_{2\rightarrow 3}(\sqrt{s}) \ .  \label{eq:P32}
\end{align}
\end{itemize}
It is important to remember that the factor ${\cal S}$ (${\cal S}'$) always refers to the number of identical particles in the initial (final) state of the particular reaction which is considered. For example the factor ${\cal S}'$ appearing in Eq.~(\ref{eq:P23bis}) is, in general, different from the one appearing in Eq.~(\ref{eq:P32}).

\vspace{5mm}

\section{${3 \rightarrow1}$ reactions \label{sec:app-31}}

In this work we have focused on ${3 \leftrightarrow2}$ reactions as they are required for deuteron production and annihilation. However, to test the stochastic criterion in \texttt{SMASH} we have also analyzed the simpler case of ${3 \leftrightarrow1}$ reactions. This was the first multi-particle reaction explored in the context of the \texttt{SMASH} approach and we report some results of this reaction here.

The $3 \rightarrow 1$ probability is applied in \texttt{SMASH} for a set of known three-body decay (back-) reactions of mesonic resonances: $\pi\pi\pi \rightarrow\omega$, $\pi\pi\pi\rightarrow\phi$ and $\pi\pi\eta\rightarrow\eta'$.  
    
In particular, for the $\omega$ and $\eta'$ the three-body decay is the dominant decay channel and with the stochastic criterion it is now possible to treat the multi-particle back-reaction directly. With the geometric criterion the back-reaction had to be treated in two steps with a resonance in the intermediate state (similar in spirit to the deuteron treatment) e.g. for the $\omega$: $\pi\pi\pi\rightarrow\rho\pi\rightarrow\omega$.

The formula for the probability is given in Eq.~(\ref{eq:P31}), where $M$ is the invariant mass of the three particles (or the total energy in the center-of-mass frame),
\be M^2 = (E_1+E_2+E_3)^2 -({\bf p}_1+ {\bf p}_2 + {\bf p}_3)^2 \ , \ee
and $\Phi_3(M^2)=\Phi_3 (M^2,m_1^2,m_2^2,m_3^2)$ is given in Eq.~(\ref{eq:Phi3}). An analytical formula for the three-body phase space is given in Ref.~\cite{Bauberger:1994by}. For the $\omega$/$\phi$ decay into three pions that formula reduces to~\cite{Davydychev:2003cw}
\begin{align} & \Phi_3 (M^2) = \frac{1}{128 \pi^3M^2} \sqrt{(M-m_\pi)(M+3m_\pi)}  \\
& \times \left[ \frac12 (M-m_\pi) (M^2+3m_\pi^2) E(k_{\textrm{eq}})-4m_\pi^2M K (k_{\textrm{eq}})] \right] \ , \nonumber
\end{align}
with the complete elliptic integrals,
\begin{align}
K(k) & = \int_0^1 \frac{dt}{\sqrt{(1-t^2)(1-k^2t^2)}} \ , \\
E(k) & = \int_0^1 dt \sqrt{ \frac{1-k^2t^2}{1-t^2}} \ , 
\end{align}
and 
\be k_{\textrm{eq}} = \sqrt{\frac{(M+m_\pi)^3(M-3m_\pi)}{(M-m_\pi)^3(M+3m_\pi)}} \ . \ee

The symmetry and degeneracy factors for $\pi \pi \pi \rightarrow \omega$ and  $\pi \pi \pi \rightarrow \phi$ reactions are
\be g_{1'}=3 \ , \quad , g_{1}=g_2=g_3=1 \ , \quad {\cal S}! = 1 \ , \ee
for $\pi^+\pi^-\eta\rightarrow\eta'$ are
\be g_{1'}=g_1=g_2=g_3=1 \ , \quad  {\cal S}!=1 \ , \ee
whereas for $\pi^0\pi^0\eta\rightarrow\eta'$ are
\be g_{1'}=g_1=g_2=g_3=1 \ , \quad  {\cal S}!=2 \ . \ee

\begin{figure}[htb]
    \centering
    \includegraphics[width=0.45\textwidth]{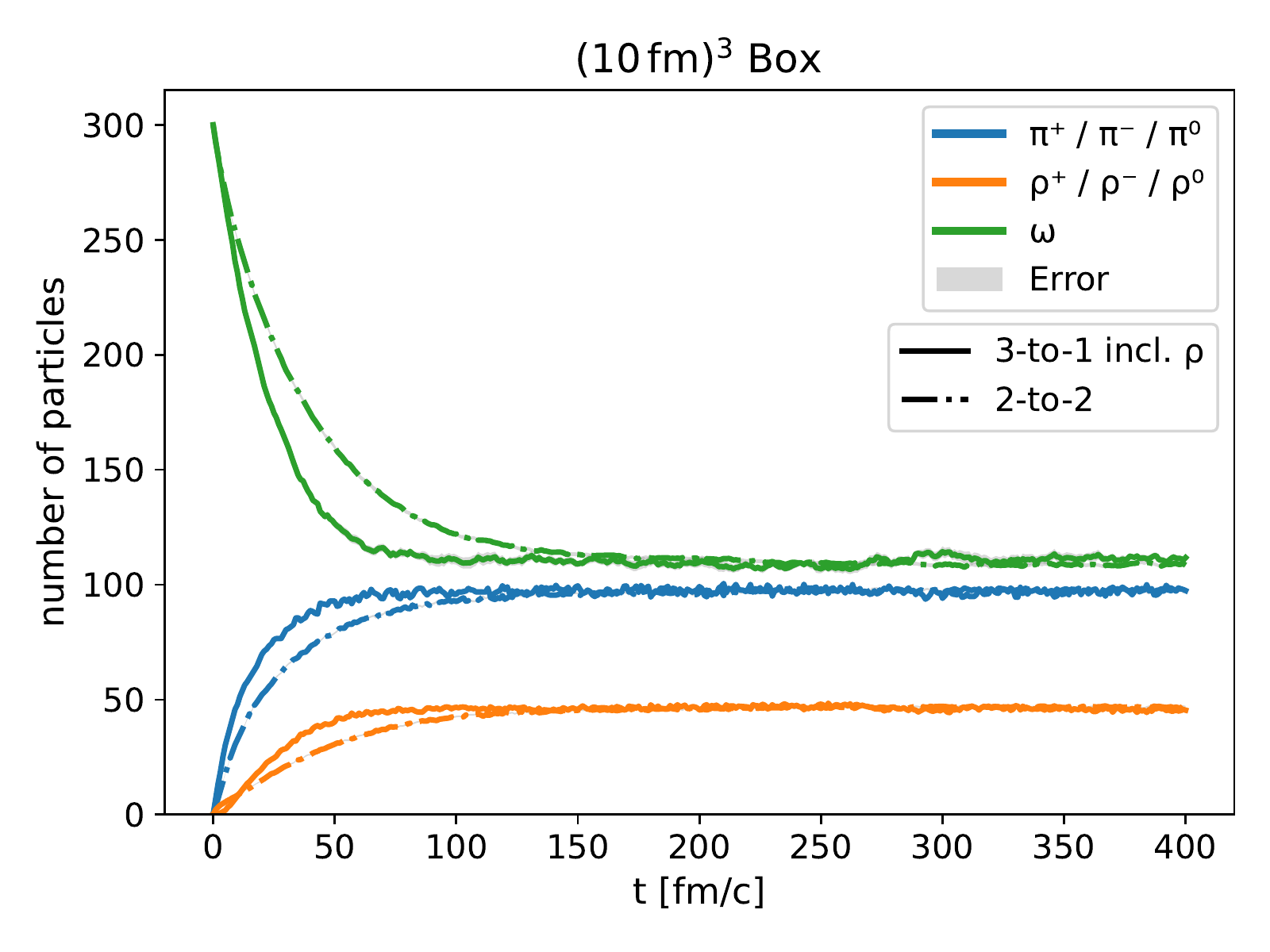}
    \caption{Evolution of particle yields in a box of $\rho, \omega$ and $\pi$.}
    \label{fig:3to1-box}
\end{figure}

While the 3-to-1 back-reaction are found to be rare in nucleus-nucleus collisions, they still allow to test and study the approach and effects for multi-particle reactions in general. Therefore, the equilibration in a box employing the reaction is checked in a similar fashion as presented above for the deuteron reactions in Figure~\ref{fig:3to1-box} for the $\pi\pi\pi \rightarrow\omega$ reaction as an example. The $\rm (10\, fm)^3$-sized box is filled with 300 $\omega$ mesons at initialization. The calculation including the 3-to-1 multi-particle reactions is displayed as the solid lines and clearly shows to equilibrate chemically.

To verify the equilibration of the system, the ratio of equilibrated particle numbers of $R_{\omega\pi}=\frac{N_\omega}{N_\pi^3}$ is compared to the thermal i.e. grand-canonical ideal gas expectation for this ratio ($R^{th}_{\omega\pi}$). For this the temperature of the box is extracted ($T= 160$ MeV) by fitting the $\pi$ energy spectrum assuming an exponential shape in equilibrium. Taking the ratio cancels the for the employed reactions specific chemical potential.

The calculated and thermal ratio are consistent with each other: $\left|\frac{R_{\omega\pi}-R_{\omega\pi}^{th}}{R_{\omega\pi}^{th}}\right| = 0.04 $, thereby verifying the here presented approach for 3-to-1 reactions. That detailed balance, in particular for the 3-to-1 reaction is fulfilled, is also checked.

Figure~\ref{fig:3to1-box} furthermore shows the results for modeling the same 3-to-1 reaction with two-body reactions ($\pi\pi\pi\rightarrow\rho\pi\rightarrow\omega$) labeled \emph{2-to-2}. Note that the $\rho$ (and the the $\rho\leftrightarrow\pi\pi$) reaction were only included in the 3-to-1 calculation to have the same degrees of freedom as in the 2-to-2 case. Its inclusion is not necessary to employ the 3-to-1 reactions of interest.
The equilibrated yield is matching for both, multi-particle and multi-step, treatment. Interestingly, comparing the two further, a very similar trend as for the same comparison for $d$ reactions is observed (cf. Figure~\ref{fig:box_rates}). The direct treatment of the 3-to-1 reaction leads to a faster equilibration of the yields. Hereby, confirming the findings made for deuterons that multi-particle reactions drive the medium faster to equilibrium.

\section{Rate equations derivation \label{sec:app-rate-equations}}

As in Ref.~\cite{Pan:2014caa} we start by assuming that all species in the box are in kinetic, but not chemical equilibrium. This means that their momentum distribution is a Boltzmann distribution at temperature $T$, but its normalization, the yield, is arbitrary. Therefore, every species $i$ is assigned a fugacity $\lambda_i$ and the multiplicity $N_i$ is written as

\begin{align}
    N_i & = V n_i^{th}(T) \lambda_i \ , \\
    n^{th}_i(T) & \equiv \frac{g_i}{(2\pi \hbar)^3} \int dM d^3p \, e^{-E_M/T} \, \mathcal{A}(M) \nonumber \\
    & = \frac{g_i T}{2\pi^2 \hbar^3} \int dM M^2 K_2(M/T) \mathcal{A}(M) \ , 
\end{align}
where $V$ is the box volume, $g_i$ is the degeneracy of the species, $\mathcal{A}(M)$ its spectral function as defined after Eq.~(\ref{eq:spectfunc}), and $K_2(x)$ is the modified Bessel function of the second kind. Finally, $E_M^2=p^2+M^2$.

We assume that the temperature is constant in time, which indeed turns out to be the case for temperature from momentum spectra in the simulations. Therefore, our rate equations are going to be equations for fugacities. The reaction rates (number of reactions per unit volume per unit time) for $2 \leftrightarrow n$ reactions indexed as $1+2 \leftrightarrow 1' + 2' + \dots + n'$ are

\begin{align}
 \frac{dN_{2\to n}}{d^4x} & = A \lambda_1 \lambda_2 \ , \\
 \frac{dN_{n\to 2}}{d^4x} & = A \prod_{j=1}^n \lambda_{j} \ , 
 \end{align}
 with
\begin{align}
 A & \equiv \int \frac{g_1 d^3p_1}{(2\pi\hbar)^3} \frac{g_2 d^3p_2}{(2\pi\hbar)^3} \, \sigma_{2\to n}  v_{\textrm{rel}} e^{-(E_1 + E_2)/T} \nn \\
 & = \langle \sigma_{2\to n} v_{\textrm{rel}} \rangle \ n^{th}_1(T) n^{th}_2(T) \ . 
\end{align}

In our case the cross section $\sigma_{2\to n}$ depends only on the center of mass energy of the reaction $\sqrt{s}$. In this case we find that $\langle \sigma_{2\to n} v_{\textrm{rel}} \rangle$ expression simplifies to~\cite{Letessier:2002gp}
\begin{align}
   \langle \sigma_{2\to n} v_{\textrm{rel}} \rangle = \frac{\int_{m_1+m_2}^{\infty} dM \sigma(M) \lambda (M^2, m_1^2, m_2^2) K_1(M/T)}{4 m_1^2 m_2^2 T K_2(m_1/T) K_2(m_2/T)} \ . 
\end{align}

This expression is equivalent to Eq. (30) from Ref.~\cite{Cannoni:2013bza}. The rate of $1 \leftrightarrow n$ decays and formations is

\begin{align}
 \frac{dN_{1\to n}}{d^4x} & = A \lambda_1 \ , \\
 \frac{dN_{n\to 1}}{d^4x} & = A \prod_{j=1}^n \lambda_{j} \ ,
\end{align}
 with
\begin{align}
A & \equiv \int dM \frac{d^3p}{(2\pi\hbar)^3 } \frac{M}{E_M} \Gamma(M)  e^{-E_M/T} \, \mathcal{A}(M) \nonumber  \\
& = \langle \Gamma \rangle \ n^{th}_1 (T) \ ,
\end{align}
where
\be
    \langle \Gamma \rangle = \frac{\int dM d^3p \ \Gamma(M) \frac{M}{E_M} e^{-E_M/T} \mathcal{A}(M)}{\int dM d^3p \ e^{-E_M/T} \mathcal{A}(M) } \ .
\ee

The factor $M/E_M$---which naturally appears in the $1\rightarrow 2$ decay probability of Eq.~(\ref{eq:P12})---can be traced back to the Lorentz time dilation. After integrating out momenta, the thermally averaged width is written as
\be    \langle \Gamma \rangle = \frac{\int dM M^2 \ K_1(M/T)\Gamma(M)\mathcal{A}(M)}{\int dM M^2 \  K_2(M/T)\mathcal{A}(M)} \ . \ee

\section{Grid cells}

As mentioned in Section~\ref{sec:st-crit}, the stochastic collision criterion divides the space into a grid with equally sized cells. The cell size $(\Delta^3x)$ is used for the collision probability calculation. Only particles within a cell interact. 
The general requirement is that the cells are small, but still sufficiently filled with (test) particles. An appropriate grid choice balances these requirements with the calculation runtime.
For the presented Au+Au afterburner calculations (Section~\ref{sec:results}), this is achieved by choosing a fixed (minimal) cell length of $l_{\rm min}=1.78 \,\rm fm$ together with a test particle number of $N_{\rm test} = 20$. The grid is chosen to be the same size as the medium i.e. all particles are inside of grid cells and updated at every timestep. The number of cells is determined by dividing $l_{\rm min}$ by the total grid length in each dimension and rounding down to an integer value. If necessary, all cell lengths are increased from their minimal value to fill the total grid length. This increase is negligible for the presented results, since the number of cells is large. 
The timestep for all calculations is $\Delta t=0.1 \,\rm fm$. 
It was separately verified that the results are stable when employing a smaller grid length, more test particles and a smaller timestep.
 
\bibliography{str_deuterons}

\providecommand{\noopsort}[1]{}\providecommand{\singleletter}[1]{#1}%
\begin{thebibliography}{42}
\expandafter\ifx\csname natexlab\endcsname\relax\def\natexlab#1{#1}\fi
\expandafter\ifx\csname bibnamefont\endcsname\relax
  \def\bibnamefont#1{#1}\fi
\expandafter\ifx\csname bibfnamefont\endcsname\relax
  \def\bibfnamefont#1{#1}\fi
\expandafter\ifx\csname citenamefont\endcsname\relax
  \def\citenamefont#1{#1}\fi
\expandafter\ifx\csname url\endcsname\relax
  \def\url#1{\texttt{#1}}\fi
\expandafter\ifx\csname urlprefix\endcsname\relax\def\urlprefix{URL }\fi
\providecommand{\bibinfo}[2]{#2}
\providecommand{\eprint}[2][]{\url{#2}}

\bibitem[{\citenamefont{Danielewicz and Bertsch}(1991)}]{Danielewicz:1991dh}
\bibinfo{author}{\bibfnamefont{P.}~\bibnamefont{Danielewicz}} \bibnamefont{and}
  \bibinfo{author}{\bibfnamefont{G.}~\bibnamefont{Bertsch}},
  \bibinfo{journal}{Nucl. Phys. A} \textbf{\bibinfo{volume}{533}},
  \bibinfo{pages}{712} (\bibinfo{year}{1991}).

\bibitem[{\citenamefont{Cassing}(2002)}]{Cassing:2001ds}
\bibinfo{author}{\bibfnamefont{W.}~\bibnamefont{Cassing}},
  \bibinfo{journal}{Nucl. Phys. A} \textbf{\bibinfo{volume}{700}},
  \bibinfo{pages}{618} (\bibinfo{year}{2002}), \eprint{nucl-th/0105069}.

\bibitem[{\citenamefont{Seifert and
  Cassing}(2018{\natexlab{a}})}]{Seifert:2017oyb}
\bibinfo{author}{\bibfnamefont{E.}~\bibnamefont{Seifert}} \bibnamefont{and}
  \bibinfo{author}{\bibfnamefont{W.}~\bibnamefont{Cassing}},
  \bibinfo{journal}{Phys. Rev. C} \textbf{\bibinfo{volume}{97}},
  \bibinfo{pages}{024913} (\bibinfo{year}{2018}{\natexlab{a}}),
  \eprint{1710.00665}.

\bibitem[{\citenamefont{Seifert and
  Cassing}(2018{\natexlab{b}})}]{Seifert:2018bwl}
\bibinfo{author}{\bibfnamefont{E.}~\bibnamefont{Seifert}} \bibnamefont{and}
  \bibinfo{author}{\bibfnamefont{W.}~\bibnamefont{Cassing}},
  \bibinfo{journal}{Phys. Rev. C} \textbf{\bibinfo{volume}{97}},
  \bibinfo{pages}{044907} (\bibinfo{year}{2018}{\natexlab{b}}),
  \eprint{1801.07557}.

\bibitem[{\citenamefont{Xu and Greiner}(2005)}]{Xu:2004mz}
\bibinfo{author}{\bibfnamefont{Z.}~\bibnamefont{Xu}} \bibnamefont{and}
  \bibinfo{author}{\bibfnamefont{C.}~\bibnamefont{Greiner}},
  \bibinfo{journal}{Phys. Rev. C} \textbf{\bibinfo{volume}{71}},
  \bibinfo{pages}{064901} (\bibinfo{year}{2005}), \eprint{hep-ph/0406278}.

\bibitem[{\citenamefont{Weil et~al.}(2016{\natexlab{a}})\citenamefont{Weil,
  Staudenmaier, and Petersen}}]{Weil:2016fxr}
\bibinfo{author}{\bibfnamefont{J.}~\bibnamefont{Weil}},
  \bibinfo{author}{\bibfnamefont{J.}~\bibnamefont{Staudenmaier}},
  \bibnamefont{and} \bibinfo{author}{\bibfnamefont{H.}~\bibnamefont{Petersen}},
  \bibinfo{journal}{J. Phys. Conf. Ser.} \textbf{\bibinfo{volume}{742}},
  \bibinfo{pages}{012034} (\bibinfo{year}{2016}{\natexlab{a}}),
  \eprint{1604.07028}.

\bibitem[{\citenamefont{Oliinychenko}(2021)}]{Oliinychenko:2020ply}
\bibinfo{author}{\bibfnamefont{D.}~\bibnamefont{Oliinychenko}},
  \bibinfo{journal}{Nucl. Phys. A} \textbf{\bibinfo{volume}{1005}},
  \bibinfo{pages}{121754} (\bibinfo{year}{2021}), \eprint{2003.05476}.

\bibitem[{\citenamefont{Oliinychenko
  et~al.}(2019{\natexlab{a}})\citenamefont{Oliinychenko, Pang, Elfner, and
  Koch}}]{Oliinychenko:2018ugs}
\bibinfo{author}{\bibfnamefont{D.}~\bibnamefont{Oliinychenko}},
  \bibinfo{author}{\bibfnamefont{L.-G.} \bibnamefont{Pang}},
  \bibinfo{author}{\bibfnamefont{H.}~\bibnamefont{Elfner}}, \bibnamefont{and}
  \bibinfo{author}{\bibfnamefont{V.}~\bibnamefont{Koch}},
  \bibinfo{journal}{Phys. Rev. C} \textbf{\bibinfo{volume}{99}},
  \bibinfo{pages}{044907} (\bibinfo{year}{2019}{\natexlab{a}}),
  \eprint{1809.03071}.

\bibitem[{\citenamefont{Oliinychenko
  et~al.}(2019{\natexlab{b}})\citenamefont{Oliinychenko, Pang, Elfner, and
  Koch}}]{Oliinychenko:2018odl}
\bibinfo{author}{\bibfnamefont{D.}~\bibnamefont{Oliinychenko}},
  \bibinfo{author}{\bibfnamefont{L.-G.} \bibnamefont{Pang}},
  \bibinfo{author}{\bibfnamefont{H.}~\bibnamefont{Elfner}}, \bibnamefont{and}
  \bibinfo{author}{\bibfnamefont{V.}~\bibnamefont{Koch}},
  \bibinfo{journal}{MDPI Proc.} \textbf{\bibinfo{volume}{10}},
  \bibinfo{pages}{6} (\bibinfo{year}{2019}{\natexlab{b}}), \eprint{1812.06225}.

\bibitem[{\citenamefont{Oliinychenko
  et~al.}(2020{\natexlab{a}})\citenamefont{Oliinychenko, Shen, and
  Koch}}]{Oliinychenko:2020znl}
\bibinfo{author}{\bibfnamefont{D.}~\bibnamefont{Oliinychenko}},
  \bibinfo{author}{\bibfnamefont{C.}~\bibnamefont{Shen}}, \bibnamefont{and}
  \bibinfo{author}{\bibfnamefont{V.}~\bibnamefont{Koch}}
  (\bibinfo{year}{2020}{\natexlab{a}}), \eprint{2009.01915}.

\bibitem[{\citenamefont{Aichelin et~al.}(2020)\citenamefont{Aichelin,
  Bratkovskaya, Le~F\`evre, Kireyeu, Kolesnikov, Leifels, Voronyuk, and
  Coci}}]{Aichelin:2019tnk}
\bibinfo{author}{\bibfnamefont{J.}~\bibnamefont{Aichelin}},
  \bibinfo{author}{\bibfnamefont{E.}~\bibnamefont{Bratkovskaya}},
  \bibinfo{author}{\bibfnamefont{A.}~\bibnamefont{Le~F\`evre}},
  \bibinfo{author}{\bibfnamefont{V.}~\bibnamefont{Kireyeu}},
  \bibinfo{author}{\bibfnamefont{V.}~\bibnamefont{Kolesnikov}},
  \bibinfo{author}{\bibfnamefont{Y.}~\bibnamefont{Leifels}},
  \bibinfo{author}{\bibfnamefont{V.}~\bibnamefont{Voronyuk}}, \bibnamefont{and}
  \bibinfo{author}{\bibfnamefont{G.}~\bibnamefont{Coci}},
  \bibinfo{journal}{Phys. Rev. C} \textbf{\bibinfo{volume}{101}},
  \bibinfo{pages}{044905} (\bibinfo{year}{2020}), \eprint{1907.03860}.

\bibitem[{\citenamefont{Cugnon}(1980)}]{Cugnon:1980zz}
\bibinfo{author}{\bibfnamefont{J.}~\bibnamefont{Cugnon}},
  \bibinfo{journal}{Phys. Rev. C} \textbf{\bibinfo{volume}{22}},
  \bibinfo{pages}{1885} (\bibinfo{year}{1980}).

\bibitem[{\citenamefont{Bass et~al.}(1998)}]{Bass:1998ca}
\bibinfo{author}{\bibfnamefont{S.}~\bibnamefont{Bass}} \bibnamefont{et~al.},
  \bibinfo{journal}{Prog. Part. Nucl. Phys.} \textbf{\bibinfo{volume}{41}},
  \bibinfo{pages}{255} (\bibinfo{year}{1998}), \eprint{nucl-th/9803035}.

\bibitem[{\citenamefont{Weil et~al.}(2016{\natexlab{b}})}]{Weil:2016zrk}
\bibinfo{author}{\bibfnamefont{J.}~\bibnamefont{Weil}} \bibnamefont{et~al.},
  \bibinfo{journal}{Phys. Rev.} \textbf{\bibinfo{volume}{C94}},
  \bibinfo{pages}{054905} (\bibinfo{year}{2016}{\natexlab{b}}),
  \eprint{1606.06642}.

\bibitem[{\citenamefont{Lang et~al.}(1993)\citenamefont{Lang, Babovsky,
  Cassing, Mosel, Reusch, and Weber}}]{LANG1993391}
\bibinfo{author}{\bibfnamefont{A.}~\bibnamefont{Lang}},
  \bibinfo{author}{\bibfnamefont{H.}~\bibnamefont{Babovsky}},
  \bibinfo{author}{\bibfnamefont{W.}~\bibnamefont{Cassing}},
  \bibinfo{author}{\bibfnamefont{U.}~\bibnamefont{Mosel}},
  \bibinfo{author}{\bibfnamefont{H.-G.} \bibnamefont{Reusch}},
  \bibnamefont{and} \bibinfo{author}{\bibfnamefont{K.}~\bibnamefont{Weber}},
  \bibinfo{journal}{Journal of Computational Physics}
  \textbf{\bibinfo{volume}{106}}, \bibinfo{pages}{391 } (\bibinfo{year}{1993}),
  ISSN \bibinfo{issn}{0021-9991},
  \urlprefix\url{http://www.sciencedirect.com/science/article/pii/S0021999183711162}.

\bibitem[{\citenamefont{Sun et~al.}(2021)\citenamefont{Sun, Wang, Ko, Ma, and
  Shen}}]{Sun:2021dlz}
\bibinfo{author}{\bibfnamefont{K.-J.} \bibnamefont{Sun}},
  \bibinfo{author}{\bibfnamefont{R.}~\bibnamefont{Wang}},
  \bibinfo{author}{\bibfnamefont{C.~M.} \bibnamefont{Ko}},
  \bibinfo{author}{\bibfnamefont{Y.-G.} \bibnamefont{Ma}}, \bibnamefont{and}
  \bibinfo{author}{\bibfnamefont{C.}~\bibnamefont{Shen}}
  (\bibinfo{year}{2021}), \eprint{2106.12742}.

\bibitem[{\citenamefont{Bratkovskaya et~al.}(2011)\citenamefont{Bratkovskaya,
  Cassing, Konchakovski, and Linnyk}}]{Bratkovskaya:2011wp}
\bibinfo{author}{\bibfnamefont{E.~L.} \bibnamefont{Bratkovskaya}},
  \bibinfo{author}{\bibfnamefont{W.}~\bibnamefont{Cassing}},
  \bibinfo{author}{\bibfnamefont{V.~P.} \bibnamefont{Konchakovski}},
  \bibnamefont{and} \bibinfo{author}{\bibfnamefont{O.}~\bibnamefont{Linnyk}},
  \bibinfo{journal}{Nucl. Phys. A} \textbf{\bibinfo{volume}{856}},
  \bibinfo{pages}{162} (\bibinfo{year}{2011}), \eprint{1101.5793}.

\bibitem[{\citenamefont{Buss et~al.}(2012)\citenamefont{Buss, Gaitanos,
  Gallmeister, van Hees, Kaskulov, Lalakulich, Larionov, Leitner, Weil, and
  Mosel}}]{Buss:2011mx}
\bibinfo{author}{\bibfnamefont{O.}~\bibnamefont{Buss}},
  \bibinfo{author}{\bibfnamefont{T.}~\bibnamefont{Gaitanos}},
  \bibinfo{author}{\bibfnamefont{K.}~\bibnamefont{Gallmeister}},
  \bibinfo{author}{\bibfnamefont{H.}~\bibnamefont{van Hees}},
  \bibinfo{author}{\bibfnamefont{M.}~\bibnamefont{Kaskulov}},
  \bibinfo{author}{\bibfnamefont{O.}~\bibnamefont{Lalakulich}},
  \bibinfo{author}{\bibfnamefont{A.}~\bibnamefont{Larionov}},
  \bibinfo{author}{\bibfnamefont{T.}~\bibnamefont{Leitner}},
  \bibinfo{author}{\bibfnamefont{J.}~\bibnamefont{Weil}}, \bibnamefont{and}
  \bibinfo{author}{\bibfnamefont{U.}~\bibnamefont{Mosel}},
  \bibinfo{journal}{Phys. Rept.} \textbf{\bibinfo{volume}{512}},
  \bibinfo{pages}{1} (\bibinfo{year}{2012}), \eprint{1106.1344}.

\bibitem[{\citenamefont{Zyla et~al.}(2020)}]{Zyla:2020zbs}
\bibinfo{author}{\bibfnamefont{P.}~\bibnamefont{Zyla}} \bibnamefont{et~al.}
  (\bibinfo{collaboration}{Particle Data Group}), \bibinfo{journal}{PTEP}
  \textbf{\bibinfo{volume}{2020}}, \bibinfo{pages}{083C01}
  (\bibinfo{year}{2020}).

\bibitem[{\citenamefont{Steinberg et~al.}(2019)\citenamefont{Steinberg,
  Staudenmaier, Oliinychenko, Li, Erkiner, and Elfner}}]{Steinberg:2018jvv}
\bibinfo{author}{\bibfnamefont{V.}~\bibnamefont{Steinberg}},
  \bibinfo{author}{\bibfnamefont{J.}~\bibnamefont{Staudenmaier}},
  \bibinfo{author}{\bibfnamefont{D.}~\bibnamefont{Oliinychenko}},
  \bibinfo{author}{\bibfnamefont{F.}~\bibnamefont{Li}},
  \bibinfo{author}{\bibfnamefont{{\"O}.}~\bibnamefont{Erkiner}},
  \bibnamefont{and} \bibinfo{author}{\bibfnamefont{H.}~\bibnamefont{Elfner}},
  \bibinfo{journal}{Phys. Rev.} \textbf{\bibinfo{volume}{C99}},
  \bibinfo{pages}{064908} (\bibinfo{year}{2019}), \eprint{1809.03828}.

\bibitem[{\citenamefont{Mohs et~al.}(2020)\citenamefont{Mohs, Ryu, and
  Elfner}}]{Mohs:2019iee}
\bibinfo{author}{\bibfnamefont{J.}~\bibnamefont{Mohs}},
  \bibinfo{author}{\bibfnamefont{S.}~\bibnamefont{Ryu}}, \bibnamefont{and}
  \bibinfo{author}{\bibfnamefont{H.}~\bibnamefont{Elfner}},
  \bibinfo{journal}{J. Phys. G} \textbf{\bibinfo{volume}{47}},
  \bibinfo{pages}{065101} (\bibinfo{year}{2020}), \eprint{1909.05586}.

\bibitem[{\citenamefont{Manley and Saleski}(1992)}]{Manley:1992yb}
\bibinfo{author}{\bibfnamefont{D.}~\bibnamefont{Manley}} \bibnamefont{and}
  \bibinfo{author}{\bibfnamefont{E.}~\bibnamefont{Saleski}},
  \bibinfo{journal}{Phys. Rev. D} \textbf{\bibinfo{volume}{45}},
  \bibinfo{pages}{4002} (\bibinfo{year}{1992}).

\bibitem[{\citenamefont{Oliinychenko
  et~al.}(2020{\natexlab{b}})\citenamefont{Oliinychenko, Steinberg, Weil,
  Staudenmaier, Kretz, Schäfer, Elfner, Ryu, Rothermel, Mohs
  et~al.}}]{dmytro_oliinychenko_2020_4336358}
\bibinfo{author}{\bibfnamefont{D.}~\bibnamefont{Oliinychenko}},
  \bibinfo{author}{\bibfnamefont{V.}~\bibnamefont{Steinberg}},
  \bibinfo{author}{\bibfnamefont{J.}~\bibnamefont{Weil}},
  \bibinfo{author}{\bibfnamefont{J.}~\bibnamefont{Staudenmaier}},
  \bibinfo{author}{\bibfnamefont{M.}~\bibnamefont{Kretz}},
  \bibinfo{author}{\bibfnamefont{A.}~\bibnamefont{Schäfer}},
  \bibinfo{author}{\bibfnamefont{H.}~\bibnamefont{Elfner}},
  \bibinfo{author}{\bibfnamefont{S.}~\bibnamefont{Ryu}},
  \bibinfo{author}{\bibfnamefont{J.}~\bibnamefont{Rothermel}},
  \bibinfo{author}{\bibfnamefont{J.}~\bibnamefont{Mohs}}, \bibnamefont{et~al.}
  (\bibinfo{year}{2020}{\natexlab{b}}),
  \urlprefix\url{https://doi.org/10.5281/zenodo.4336358}.

\bibitem[{\citenamefont{Hirano and Nara}(2012)}]{Hirano:2012yy}
\bibinfo{author}{\bibfnamefont{T.}~\bibnamefont{Hirano}} \bibnamefont{and}
  \bibinfo{author}{\bibfnamefont{Y.}~\bibnamefont{Nara}},
  \bibinfo{journal}{PTEP} \textbf{\bibinfo{volume}{2012}},
  \bibinfo{pages}{01A203} (\bibinfo{year}{2012}), \eprint{1203.4418}.

\bibitem[{\citenamefont{Oh et~al.}(2009)\citenamefont{Oh, Lin, and
  Ko}}]{Oh:2009gx}
\bibinfo{author}{\bibfnamefont{Y.}~\bibnamefont{Oh}},
  \bibinfo{author}{\bibfnamefont{Z.-W.} \bibnamefont{Lin}}, \bibnamefont{and}
  \bibinfo{author}{\bibfnamefont{C.~M.} \bibnamefont{Ko}},
  \bibinfo{journal}{Phys. Rev. C} \textbf{\bibinfo{volume}{80}},
  \bibinfo{pages}{064902} (\bibinfo{year}{2009}), \eprint{0910.1977}.

\bibitem[{\citenamefont{Longacre}(2013)}]{Longacre:2013apa}
\bibinfo{author}{\bibfnamefont{R.~S.} \bibnamefont{Longacre}}
  (\bibinfo{year}{2013}), \eprint{1311.3609}.

\bibitem[{\citenamefont{Kodama et~al.}(1984)\citenamefont{Kodama, Duarte,
  Chung, Donangelo, and Nazareth}}]{Kodama:1983yk}
\bibinfo{author}{\bibfnamefont{T.}~\bibnamefont{Kodama}},
  \bibinfo{author}{\bibfnamefont{S.}~\bibnamefont{Duarte}},
  \bibinfo{author}{\bibfnamefont{K.}~\bibnamefont{Chung}},
  \bibinfo{author}{\bibfnamefont{R.}~\bibnamefont{Donangelo}},
  \bibnamefont{and} \bibinfo{author}{\bibfnamefont{R.}~\bibnamefont{Nazareth}},
  \bibinfo{journal}{Phys. Rev. C} \textbf{\bibinfo{volume}{29}},
  \bibinfo{pages}{2146} (\bibinfo{year}{1984}).

\bibitem[{\citenamefont{Schenke et~al.}(2010)\citenamefont{Schenke, Jeon, and
  Gale}}]{Schenke:2010nt}
\bibinfo{author}{\bibfnamefont{B.}~\bibnamefont{Schenke}},
  \bibinfo{author}{\bibfnamefont{S.}~\bibnamefont{Jeon}}, \bibnamefont{and}
  \bibinfo{author}{\bibfnamefont{C.}~\bibnamefont{Gale}},
  \bibinfo{journal}{Phys. Rev. C} \textbf{\bibinfo{volume}{82}},
  \bibinfo{pages}{014903} (\bibinfo{year}{2010}), \eprint{1004.1408}.

\bibitem[{\citenamefont{Schenke et~al.}(2012)\citenamefont{Schenke, Jeon, and
  Gale}}]{Schenke:2011bn}
\bibinfo{author}{\bibfnamefont{B.}~\bibnamefont{Schenke}},
  \bibinfo{author}{\bibfnamefont{S.}~\bibnamefont{Jeon}}, \bibnamefont{and}
  \bibinfo{author}{\bibfnamefont{C.}~\bibnamefont{Gale}},
  \bibinfo{journal}{Phys. Rev. C} \textbf{\bibinfo{volume}{85}},
  \bibinfo{pages}{024901} (\bibinfo{year}{2012}), \eprint{1109.6289}.

\bibitem[{\citenamefont{Paquet et~al.}(2016)\citenamefont{Paquet, Shen,
  Denicol, Luzum, Schenke, Jeon, and Gale}}]{Paquet:2015lta}
\bibinfo{author}{\bibfnamefont{J.-F.} \bibnamefont{Paquet}},
  \bibinfo{author}{\bibfnamefont{C.}~\bibnamefont{Shen}},
  \bibinfo{author}{\bibfnamefont{G.~S.} \bibnamefont{Denicol}},
  \bibinfo{author}{\bibfnamefont{M.}~\bibnamefont{Luzum}},
  \bibinfo{author}{\bibfnamefont{B.}~\bibnamefont{Schenke}},
  \bibinfo{author}{\bibfnamefont{S.}~\bibnamefont{Jeon}}, \bibnamefont{and}
  \bibinfo{author}{\bibfnamefont{C.}~\bibnamefont{Gale}},
  \bibinfo{journal}{Phys. Rev. C} \textbf{\bibinfo{volume}{93}},
  \bibinfo{pages}{044906} (\bibinfo{year}{2016}), \eprint{1509.06738}.

\bibitem[{\citenamefont{Denicol et~al.}(2018)\citenamefont{Denicol, Gale, Jeon,
  Monnai, Schenke, and Shen}}]{Denicol:2018wdp}
\bibinfo{author}{\bibfnamefont{G.~S.} \bibnamefont{Denicol}},
  \bibinfo{author}{\bibfnamefont{C.}~\bibnamefont{Gale}},
  \bibinfo{author}{\bibfnamefont{S.}~\bibnamefont{Jeon}},
  \bibinfo{author}{\bibfnamefont{A.}~\bibnamefont{Monnai}},
  \bibinfo{author}{\bibfnamefont{B.}~\bibnamefont{Schenke}}, \bibnamefont{and}
  \bibinfo{author}{\bibfnamefont{C.}~\bibnamefont{Shen}},
  \bibinfo{journal}{Phys. Rev. C} \textbf{\bibinfo{volume}{98}},
  \bibinfo{pages}{034916} (\bibinfo{year}{2018}), \eprint{1804.10557}.

\bibitem[{\citenamefont{Shen and Alzhrani}(2020)}]{Shen:2020jwv}
\bibinfo{author}{\bibfnamefont{C.}~\bibnamefont{Shen}} \bibnamefont{and}
  \bibinfo{author}{\bibfnamefont{S.}~\bibnamefont{Alzhrani}},
  \bibinfo{journal}{Phys. Rev. C} \textbf{\bibinfo{volume}{102}},
  \bibinfo{pages}{014909} (\bibinfo{year}{2020}), \eprint{2003.05852}.

\bibitem[{\citenamefont{Monnai et~al.}(2019)\citenamefont{Monnai, Schenke, and
  Shen}}]{Monnai:2019hkn}
\bibinfo{author}{\bibfnamefont{A.}~\bibnamefont{Monnai}},
  \bibinfo{author}{\bibfnamefont{B.}~\bibnamefont{Schenke}}, \bibnamefont{and}
  \bibinfo{author}{\bibfnamefont{C.}~\bibnamefont{Shen}},
  \bibinfo{journal}{Phys. Rev. C} \textbf{\bibinfo{volume}{100}},
  \bibinfo{pages}{024907} (\bibinfo{year}{2019}), \eprint{1902.05095}.

\bibitem[{\citenamefont{Adamczyk et~al.}(2017)}]{Adamczyk:2017iwn}
\bibinfo{author}{\bibfnamefont{L.}~\bibnamefont{Adamczyk}} \bibnamefont{et~al.}
  (\bibinfo{collaboration}{STAR}), \bibinfo{journal}{Phys. Rev. C}
  \textbf{\bibinfo{volume}{96}}, \bibinfo{pages}{044904}
  (\bibinfo{year}{2017}), \eprint{1701.07065}.

\bibitem[{\citenamefont{Pan and Pratt}(2014)}]{Pan:2014caa}
\bibinfo{author}{\bibfnamefont{Y.}~\bibnamefont{Pan}} \bibnamefont{and}
  \bibinfo{author}{\bibfnamefont{S.}~\bibnamefont{Pratt}},
  \bibinfo{journal}{Phys. Rev. C} \textbf{\bibinfo{volume}{89}},
  \bibinfo{pages}{044911} (\bibinfo{year}{2014}).

\bibitem[{\citenamefont{Adam et~al.}(2019)}]{Adam:2019wnb}
\bibinfo{author}{\bibfnamefont{J.}~\bibnamefont{Adam}} \bibnamefont{et~al.}
  (\bibinfo{collaboration}{STAR}), \bibinfo{journal}{Phys. Rev. C}
  \textbf{\bibinfo{volume}{99}}, \bibinfo{pages}{064905}
  (\bibinfo{year}{2019}), \eprint{1903.11778}.

\bibitem[{\citenamefont{Anticic et~al.}(2016)}]{Anticic:2016ckv}
\bibinfo{author}{\bibfnamefont{T.}~\bibnamefont{Anticic}} \bibnamefont{et~al.}
  (\bibinfo{collaboration}{NA49}), \bibinfo{journal}{Phys. Rev. C}
  \textbf{\bibinfo{volume}{94}}, \bibinfo{pages}{044906}
  (\bibinfo{year}{2016}), \eprint{1606.04234}.

\bibitem[{\citenamefont{Adamczyk et~al.}(2016)}]{Adamczyk:2016gfs}
\bibinfo{author}{\bibfnamefont{L.}~\bibnamefont{Adamczyk}} \bibnamefont{et~al.}
  (\bibinfo{collaboration}{STAR}), \bibinfo{journal}{Phys. Rev. C}
  \textbf{\bibinfo{volume}{94}}, \bibinfo{pages}{034908}
  (\bibinfo{year}{2016}), \eprint{1601.07052}.

\bibitem[{\citenamefont{Bauberger et~al.}(1995)\citenamefont{Bauberger,
  Berends, Bohm, and Buza}}]{Bauberger:1994by}
\bibinfo{author}{\bibfnamefont{S.}~\bibnamefont{Bauberger}},
  \bibinfo{author}{\bibfnamefont{F.~A.} \bibnamefont{Berends}},
  \bibinfo{author}{\bibfnamefont{M.}~\bibnamefont{Bohm}}, \bibnamefont{and}
  \bibinfo{author}{\bibfnamefont{M.}~\bibnamefont{Buza}},
  \bibinfo{journal}{Nucl. Phys. B} \textbf{\bibinfo{volume}{434}},
  \bibinfo{pages}{383} (\bibinfo{year}{1995}), \eprint{hep-ph/9409388}.

\bibitem[{\citenamefont{Davydychev and Delbourgo}(2004)}]{Davydychev:2003cw}
\bibinfo{author}{\bibfnamefont{A.~I.} \bibnamefont{Davydychev}}
  \bibnamefont{and}
  \bibinfo{author}{\bibfnamefont{R.}~\bibnamefont{Delbourgo}},
  \bibinfo{journal}{J. Phys. A} \textbf{\bibinfo{volume}{37}},
  \bibinfo{pages}{4871} (\bibinfo{year}{2004}), \eprint{hep-th/0311075}.

\bibitem[{\citenamefont{Letessier and Rafelski}(2002)}]{Letessier:2002gp}
\bibinfo{author}{\bibfnamefont{J.}~\bibnamefont{Letessier}} \bibnamefont{and}
  \bibinfo{author}{\bibfnamefont{J.}~\bibnamefont{Rafelski}},
  \emph{\bibinfo{title}{{Hadrons and quark - gluon plasma}}}
  (\bibinfo{publisher}{Cambridge University Press}, \bibinfo{year}{2002}), ISBN
  \bibinfo{isbn}{978-0-521-01823-4, 978-0-521-38536-7, 978-0-511-03727-6}.

\bibitem[{\citenamefont{Cannoni}(2014)}]{Cannoni:2013bza}
\bibinfo{author}{\bibfnamefont{M.}~\bibnamefont{Cannoni}},
  \bibinfo{journal}{Phys. Rev. D} \textbf{\bibinfo{volume}{89}},
  \bibinfo{pages}{103533} (\bibinfo{year}{2014}), \eprint{1311.4494}.

\end{thebibliography}

\end{document}